\documentclass[
aip,
amsmath,amssymb,
preprint,floatfix
]{revtex4-1}

\usepackage{physics}
\usepackage{verbatim}
\usepackage{natbib}
\usepackage{placeins}
\usepackage{adjustbox}
\usepackage{soul}
\usepackage{subcaption}
\usepackage{cleveref}
\usepackage{graphicx}
\usepackage{float}
\usepackage{xcolor}

\Crefname{equation}{Eq.}{Eqs.}
\Crefname{figure}{Fig.}{Figs.}
\DeclareMathAlphabet{\mathpzc}{OT1}{pzc}{m}{it}
\DeclareMathAlphabet\mathbfcal{OMS}{cmsy}{b}{n}
\begin{document}

	\title{Backward waves in the nonlinear regime of the Buneman instability}
	
	\author{Arash Tavassoli}
	\email{art562@mail.usask.ca}
	\author{Magdi Shoucri}
	\author{Andrei Smolyakov}
	\author{Mina Papahn Zadeh}
	\affiliation{Department of Physics and Engineering Physics, University of Saskatchewan,  Saskatoon,  Saskatchewan, Canada S7N 5E2}
	\author{Raymond J. Spiteri}
	\affiliation{Department of Computer Science, University of Saskatchewan,  Saskatoon, Saskatchewan, Canada S7N 5C9}
	\date{\today}
	\begin{abstract}
	    Observation of low- and high-frequency backward waves in the nonlinear regime of the Buneman instability is reported. Intense low-frequency backward waves propagating in the direction opposite to the electron drift (with respect to the ion population) of ions and electrons are found.  The excitation of these waves is explained based on the linear 
	   theory for the stability of the electron velocity distribution function that is modified by nonlinear effects. In the nonlinear regime, the electron distribution exhibits a wide plateau formed by electron hole trapping and extends into the negative velocity region. It is shown that within the linear approach, the backward waves correspond to the weakly unstable or marginally stable modes generated by the large population of particles with negative velocities.
	\end{abstract}
	\maketitle

\section{Introduction}	
The Buneman instability is a two-stream type instability driven by the relative drift $v_0$ of electrons with respect to ions in an unmagnetized cold plasma.
It has been studied in numerous settings as a mechanism of turbulence and source of anomalous resistivity in space plasmas \cite{che2009nonlinear,CheHH_MPLA2016,HellingerGRL2004,che2010electron,DyrudJGR2006}, as a generation mechanism for  short wavelength radiation sources \cite{CarlstenPoP2008}, in ion beam fusion applications \cite{StartsevPoP2006}, and many others.
The linear regime of this instability has been well studied and understood for some time \cite{buneman1958instability,buneman1959dissipation}. 
In contrast, the nonlinear regime is complicated, and its various aspects are still subjects of interest. The nonlinear dynamics of trapping and the resultant holes\cite{hutchinson2017electron,califano2007electrostatic,ghizzo1988nonlinear,shoucri2017formation}, the long-time behaviour of the nonlinear regime\cite{ghizzo1988stability,manfredi1997long,knorr1977two}, and nonlinear Landau damping\cite{villani2014particle} are among such aspects of the problem. In this regard, numerical simulations play an important role complementing the analytical theories. Over the past decades, many numerical studies have been performed to reveal various nonlinear phenomena in the Buneman  instability\cite{rajawat2017particle,BuchnerPoP2006,lampe1974two,shokri2005nonlinear,niknam2011simulation,niknam2011simulation,HaraPoP2018}. Despite these efforts, however, theoretical explanations for a variety of the observed nonlinear phenomena remain elusive. 

Ref.~\onlinecite{bartlett1968nonadiabatic} provides one of the first descriptions of the effects of nonlinear mode coupling. It predicts a decline in the linear growth rate accompanied by a nonlinear, oscillatory growth, but it fails to predict the saturation level of the instability. Following a similar approach, Refs.~\onlinecite{ishihara1980nonlinear,ishihara1981nonlinear} calculate the ion susceptibility, taking into account the nonlinear mode-coupling and thus expanding the quasi-linear dispersion relation into the nonlinear regime. In contrast with Ref.~\onlinecite{bartlett1968nonadiabatic}, this theory  predicts the saturation, the initial depression of the relative drift velocity, and the initial heating of electrons. However, as soon as electron trapping becomes important, this theory fails in its predictions for various quantities such as drift velocity and ion susceptibility. Electron trapping is later incorporated into the model in a companion paper \cite{hirose1982nonlinear}, and the effects on the ion dynamics are investigated. In another effort, Ref.~\onlinecite{yoon2010weak} develops a weak turbulence theory for the nonlinear regime of the Buneman instability. This theory is shown \cite{yoon2010nonlinear} to explain some characteristics of the nonlinear evolution seen in numerical simulations. However, the shape of the electron velocity distribution function (VDF) is taken as a shifted Maxwellian at all times, whereas various simulations show that in the nonlinear regime, the VDF deviates significantly from a shifted Maxwellian. The criteria for applicability of quasilinear theory are not satisfied  in many situations\cite{SigovPPCF1996}. It is our goal in this study to investigate the nonlinear stage of the strong Buneman instability when the whole electron population is streaming with respect to the ions,  both components are warm, and they have the same initial temperatures. This regime is characterized by the excitation of large-amplitude fluctuations of the potential, the strong modification of the electron distribution function and heating due to the electron reflections and trapping, and as a consequence,  the excitation  of the waves in the direction opposite to the beam velocity.   

\begin{figure}[htbp]
\centering
\includegraphics[width=.49\linewidth]{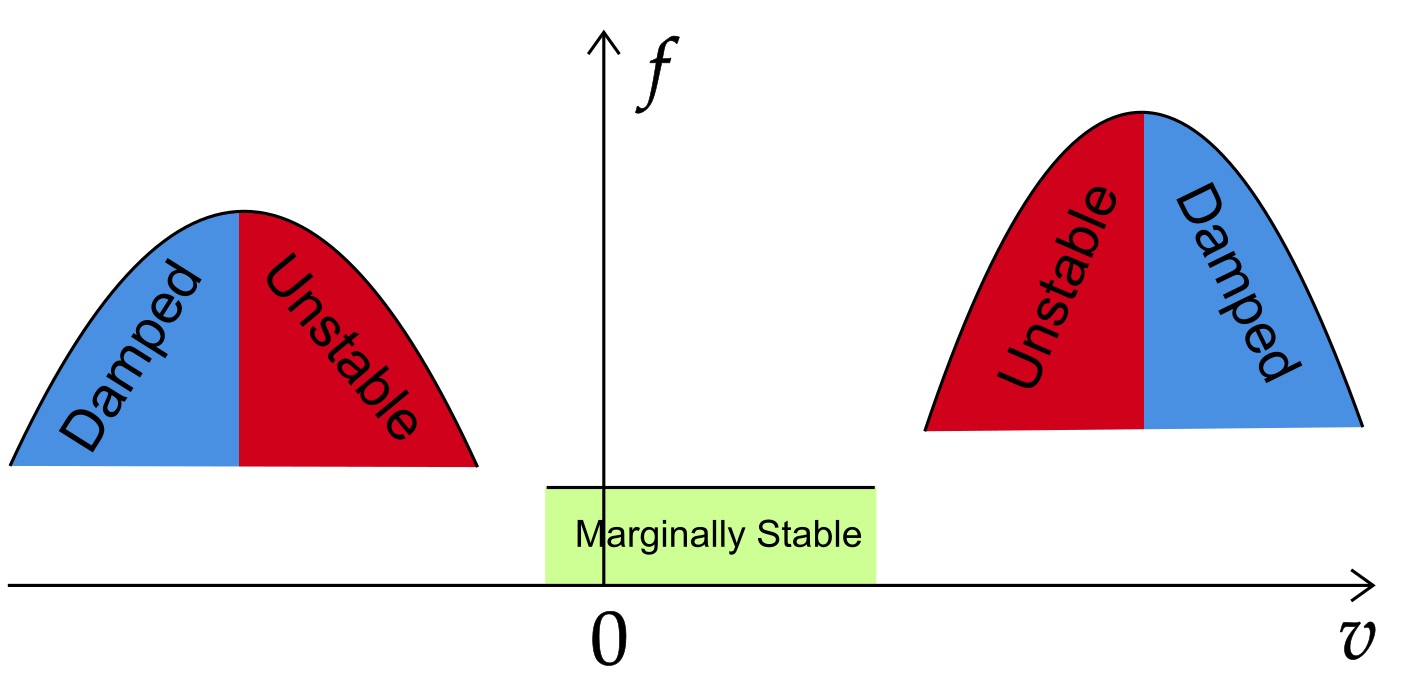}
\caption{The regions of damped, unstable, and marginally stable modes according to the local criterion from the  linear theory of Landau damping.}
\label{Landau_damping}
\end{figure}

In many situations, large deviation of the distribution function from the initial Maxwellian is a defining feature of the nonlinear evolution. This feature has led to an approach in which the nonlinearly modified VDF is used in the linear dispersion relation to interpret and explain the mode behavior. For example, the suppression of Landau damping in the nonlinear regime can be understood from the fact that the nonlinear VDF develops a plateau that stops the Landau damping \cite{mazitov1965damping} due to the commonly used local criterion  $\partial f/ \partial v >0$ from the  linear theory of Landau damping (\Cref{Landau_damping}). This criterion suggests that for $\partial f/ \partial v >0$, the modes with phase velocity close to the resonant condition $\omega =k v $  become unstable, whereas in  the region of the negative slope of the VDF with  $\partial f/ \partial v <0$, the modes are damped. It is important to note that for the waves with negative phase velocity and for which the velocity of resonant particles
is also negative, $v=\omega /k<0$, the situation  is reversed, so that $\partial f/ \partial v <0$ is required for the instability, and  for $\partial f/ \partial v >0$, resonant modes are damped. In both cases, a region of zero slope in the VDF (or a plateau) leads to the marginal stability of the modes with the phase velocity  in the plateau region; see \Cref{Landau_damping}. 

Although the local instability criterion is often useful and  insightful,   the whole profile of the distribution function is required in general to determine the full linear stability as embodied in various  integral stability criteria, e.g., the Penrose criterion \cite{krall1973principles}. 
The linear analysis   on the modified distribution function  has explained new types of  waves  such as nonlinear electron-acoustic waves (EAWs)   \cite{valentini2006excitation,valentini2012undamped,valentini2013response}.  EAWs are a class of nonlinear waves with phase velocity close to the electron thermal velocity. According to the linear theory of Landau damping, these modes are expected to be damped, whereas simulations show that they are marginally stable. To explain this behavior, the theory developed in  Refs.~\onlinecite{valentini2012undamped} and \onlinecite{holloway1991undamped} used the standard  Maxwellian VDF with an additional term that accounts for the small plateau seen in the simulations. Based on the linear dispersion relation, this plateau is found to be responsible for new modes termed ``corner modes". The existence of a large-amplitude wave, propagating with the  phase velocity near the zero-slope inflection point of the electron VDF, has also been demonstrated in Ref.~\onlinecite{shoucri2017formation}. Even though Ref.~\onlinecite{schamel2013comment}  argued that a purely nonlinear theory taking into account particle trapping is necessary to explain the new modes seen in Ref.~\onlinecite{valentini2012undamped}, it seems unavoidable that the existence of these modes is in  large part  related  to the suppression of Landau damping due to flattening of the distribution function, an effect that has clear interpretation in the linear theory. 

The linear analysis using the dispersion relation based on the modified distribution function is also used in other nonlinear studies \cite{che2009nonlinear,che2010electron,pavan2011quasilinear,jain2011modeling}. In Refs.~\onlinecite{che2009nonlinear,che2010electron,pavan2011quasilinear,jain2011modeling}, a two-Maxwellian VDF was fit to the nonlinear VDF found from the simulations. The resulting distribution function is then fed into the linear dispersion relation. The solution of this dispersion relation leads to the frequencies and growth rates of the modes observed in the simulations. 

We employ a similar approach in this study, where using a low-noise, high-resolution, grid-based Vlasov solver, we observe  the nonlinear excitation of strong waves that propagate in the direction opposite to that of the initial electron drift. These waves are referred to below as \textit{backward waves}. Backward waves have been observed in some simulations of current-driven instabilities, and they are believed to have important effects on the nonlinear evolution of these instabilities
 \cite{HaraPSST2019,HellingerGRL2004,jun2012competition,jain2011modeling,DyrudJGR2006}. However, the origin of these backward waves is still not well understood. Different scenarios such as  secondary linear instability \cite{jain2011modeling},  three-wave decay\cite{YoonPoP2018}, and  induced scattering off ions\cite{HellingerGRL2004,YoonPoP2018} are mentioned  as possible mechanisms for the excitation of these waves. The backward waves are generated well into the nonlinear regime and are not generally expected based on the linear theory of the Buneman instability, in which the Maxwellian electron population is streaming with respect to the (also Maxwellian) ions. In this study, we report  the observations of low-frequency (ion-sound-like) and high-frequency (Langmuir-like) backward waves. The intensity of the ion-sound-like waves is much higher than that of the high-frequency mode, the amplitude of which  also decreases further into the nonlinear regime.

 In the nonlinear stage, the electron VDF strongly deviates from  Maxwellian. The  electron VDF is  modified by the trapping in the holes and forms a plateau that extends well into the negative velocity region. The plateau in the negative velocity region allows for the existence of weakly unstable and marginally stable backward (and forward) waves that otherwise would suffer Landau damping. This situation is similar to the observations in Refs.~\onlinecite{valentini2012undamped,valentini2011new,shoucri2017formation,johnston2009persistent}, where the trapping of electrons or ions forms a plateau in the VDF and allows for a new class of waves. Here, we show that the nonlinear VDF observed in simulations is susceptible to the excitation of backward waves observed in  simulations.   

 We use the VDF from simulations averaged over time intervals of about $15\; \omega_{pi}^{-1} \sim 20\; \omega_{pi}^{-1}$, where $\omega_{pi}$ is the ion plasma frequency.   These time intervals are long enough to get  clear view of the low-frequency modes in the fast Fourier transform (FFT) of the electric field from nonlinear simulations.  
 These spectra are compared with results of the linear stability analysis performed on the actual distribution function obtained  by averaging  for each interval.
 
The remainder of the paper is organized as follows. In \cref{LinearTheory}, we review the linear dispersion relation of the Buneman instability and its solution for the cases of this study. In particular, we show that the linear growth rates calculated  our simulations are in agreement with the linear theory for the initial  Maxwellian distribution. In \cref{ProbSet}, the general setup of the nonlinear problem and simulations is reported. We have performed simulations with $v_0=4v_{te}$ and $v_0=10v_{te}$. In \cref{analysis}, the linear analysis based on the modified VDF is applied to the simulation results, and its predictions with regard to backward and forward waves are discussed. In \cref{Threshold}, two other simulations with  $v_0=1.5v_{te}$ and $v_0=1.75v_{te}$ are compared, and accordingly, we show that a threshold for excitation of backward waves lies between these two values. We also show that for values of $v_0$ less than this threshold, the formed plateau in electron VDF is not wide enough to extend into the negative velocity region, and therefore, backward waves do not appear. In \cref{discussion}, we conclude our study with a discussion of the results.

\section{Linear regime of the Buneman instability}\label{LinearTheory}

The Buneman instability is the electrostatic  instability driven by the relative drift of plasma species. In the limit that both electron and ion temperatures vanish, the dispersion relation of the Buneman instability is
\begin{align}
    1-\frac{\omega_{pe}^2}{(\omega-kv_0)^2}-\frac{\omega_{pi}^2}{\omega^2}=0.
\end{align}
Here,  $\omega$ is the eigen-mode frequency, $k$ is the wave vector, $v_0$ is the initial drift velocity of the electrons, $\omega_{pi}$ is the ion plasma frequency, and  $\omega_{pe}$ is the electron plasma frequency. The instability occurs for $k v_0 <\omega_{pe} (1+(m_e/m_i)^{1/3})^{3/2}$, with the  maximum mode growth rate $\gamma=\frac{\sqrt{3}}{2}(\frac{m_e}{2m_i})^{1/3} \omega_{pe}$ at  $k\approx \omega_{pe}/v_0$, and real part of the frequency $\omega=\frac{1}{2}(\frac{m_e}{2m_i})^{1/3}\omega_{pe}$.   

Considering ions and electrons with finite temperatures, the dispersion relation reads
\begin{gather}
    1-\frac{\omega_{pi}^2}{2k^2v_{ti}^2}Z^{'}\qty(\frac{\omega}{\sqrt{2}\abs{k}v_{ti}})-\frac{\omega_{pe}^2}{2k^2v_{te}^2}Z^{'}\qty(\frac{\omega-kv_{0}}{\sqrt{2}\abs{k}v_{te}})
    =0,
    \label{disper_linear}
\end{gather}
where $v_{ti}=\sqrt{T_{i0}/m_i}$ and $v_{te}=\sqrt{T_{e0}/m_e}$ are the ion and electron initial thermal velocities and $T_{i0}$ and $T_{e0}$ are the initial temperature of ions and electrons. In this study, we take $T_0=0.2$ eV as the initial temperature for both ions and electrons. The equal ion and electron temperature regime is of particular interest for the study of solar plasmas\cite{CheHH_MPLA2016}. We note that therefore, the ion sound velocity $c_s$ is equal to the $v_{ti}$. Also, we take $n_0=10^{17}\text{m}^{-3}$ as the plasma density. 

\begin{figure}[htbp]
\centering
\captionsetup[subfigure]{labelformat=empty}
\subcaptionbox{\label{linear_4vte}}{\includegraphics[width=.49\linewidth]{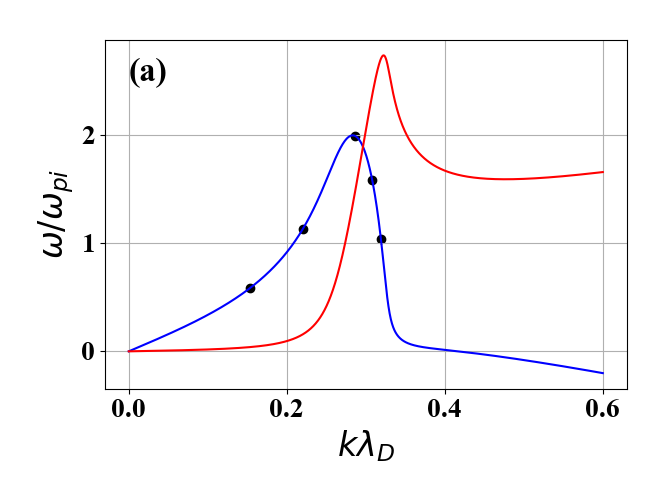}}
\subcaptionbox{\label{linear_10vte}}{\includegraphics[width=.49\linewidth]{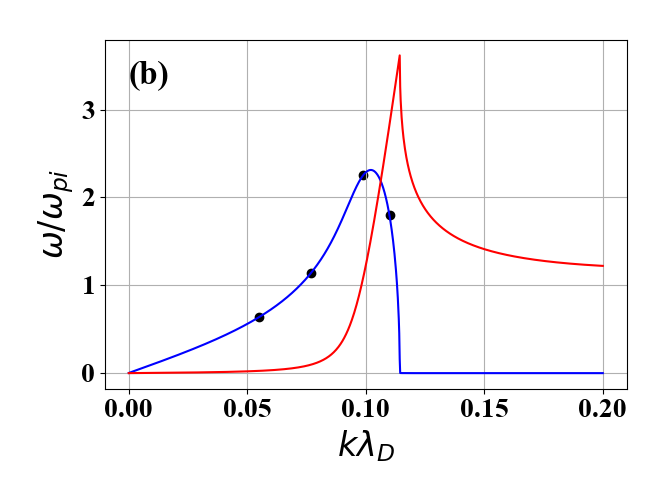}}
\caption{(a) The linear growth rate (blue) and frequency (red) for the case of $v_0=4v_{te}$. (b) The linear growth rate (blue) and frequency (red) for the case of $v_0=10v_{te}$. The black circles show the growth rates measured from the simulation. (Note the different axis ranges.)}
\end{figure}

Using these parameters, \Cref{linear_4vte,linear_10vte} show the solution of \Cref{disper_linear}, for the two drift velocities cases $v_0=4v_{te}$ and $v_0=10v_{te}$ (note the different axis ranges). These two drift velocities are also considered in the nonlinear simulations of this study. With periodic boundary conditions, as used here, only the modes satisfying the condition $kL/{2\pi}=m$ for integer $m$ are allowed.  
In the case of $v_0=4v_{te}$, the most unstable mode corresponds the wave number $m=26$. The case $v_0=10v_{te}$ is closer to the cold plasma limit $v_0/v_{te}\gg 1$. In this case, the positive growth rate region is shorter, and the most unstable mode corresponds to the wave number $m=9$. The growth rate diagram is also sharper, and the maximum growth rate has increased. Backward waves cannot be observed in the solution of linear dispersion of the Buneman instability, and therefore, we have omitted the negative $k$ region from \Cref{linear_4vte,linear_10vte}. In \Cref{linear_4vte,linear_10vte}, we have also reported the growth rate of some chosen electric field modes as measured from the simulations. The evolution of the amplitude of these chosen modes is shown in \Cref{t_Ek_v04,t_Ek_v010}. As we see, after some initial oscillations, the amplitude of each mode shows linear growth. The slope of this linear growth is used to measure the growth rate of each mode.

\section{Nonlinear Vlasov simulations}\label{ProbSet}

In our simulations, we solve the Vlasov--Poisson equations
\begin{align}
\pdv{f_{\ell}}{t}+v_x\pdv{f_{\ell}}{x}+ \frac{qE_x}{m_{\ell}}\pdv{f_{\ell}}{v_x}&=0,\nonumber\\
\pdv{E_x}{x}&=e(n_i-n_e),
\end{align}
where $f_{\ell}$ is the distribution function for species $\ell$, $\ell = i, e$ for ions and electrons, respectively, $E_x$ is the electric field, $n_{\ell}=\int f_{\ell}\,\dd v_x$ is the density of species $\ell$, $q$ is the charge, which is $+e$ for the ions and $-e$ for the electrons, and $m_{\ell}$ is the mass of species $\ell$. The ions are taken to be Hydrogen with mass $m_i=1$ amu.

The numerical method used is the well-known and tested semi-Lagrangian splitting scheme \cite{gagne1977splitting,cheng1976integration}. In this method, the Vlasov equation is split into a convection equation and a force equation. Each of these equations is then solved by the method of characteristics and cubic spline interpolation. The boundary condition is periodic in space and open in the velocity direction. The Poisson equation is solved by a spectral method, the FFT. The initial conditions are
\begin{align}
f_e(x,v,0)&=\frac{n_0(1+\epsilon k_0 \cos(k_0x))}{\sqrt{2\pi}v_{te}}\exp(-\frac{(v-v_0)^2}{2v_{te}^2}), \\
f_i(x,v,0)&=\frac{n_0}{\sqrt{2\pi}v_{ti}}\exp(-\frac{v^2}{2v_{ti}^2}).
\end{align}
 The quantities $\epsilon=10^{-8} \lambda_D$, where $\lambda_D\equiv \sqrt{\epsilon_0T_0/n_0e^2}$ is the Debye length, and $k_0=\frac{2\pi}{L}$ parameterize an initial small perturbation. These parameters are required to excite the instability because the method for solving the Vlasov formulation inherently introduces little numerical noise. For this study, we also tried the perturbation with the most unstable mode and also several perturbed modes, and we confirmed that the results are not sensitive to the choice of initial perturbation. The system length is taken $L=6$ mm, which is approximately $570$ Debye lengths, and a spatial grid of 4096 points is used. This length is large enough to contain many unstable modes, including the mode with maximum linear growth rate (see \Cref{linear_4vte,linear_10vte}). The velocity grids for ions and electrons consist of 1921 and 4033 points, respectively. The time-step used in the simulations is $2\times10^{-3}\;\omega_{pi}^{-1}$, which is about $0.086\;\omega_{pe}^{-1}$. Therefore, the time step is small enough to resolve fast variations of the plasma. We note that $\omega_{pi}^{-1}=2.39$ ns in this setup.

\begin{figure}[htbp]
\centering
\captionsetup[subfigure]{labelformat=empty}
\subcaptionbox{\label{t_efE_v04}}{\includegraphics[width=.49\linewidth]{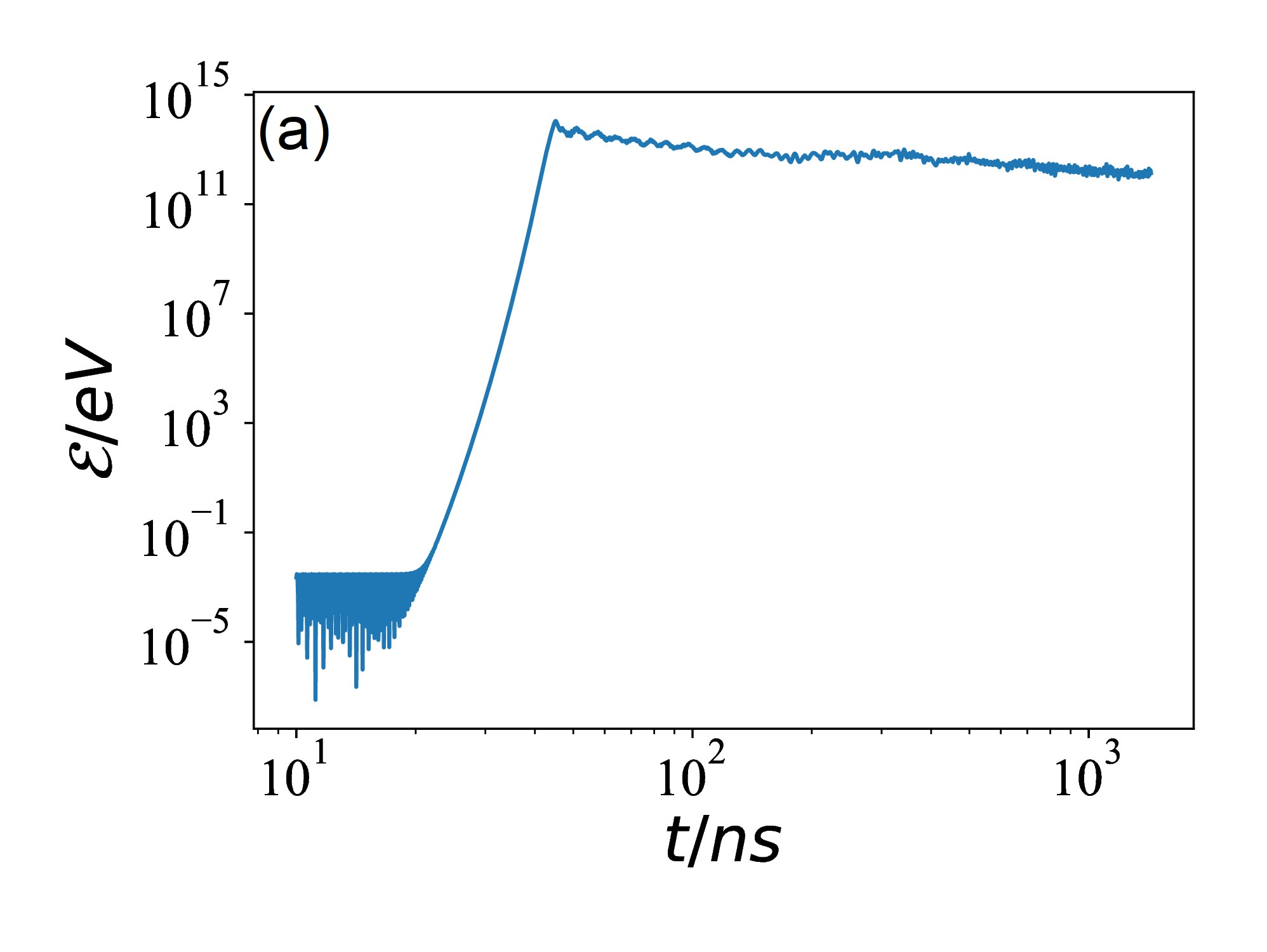}}
\subcaptionbox{\label{t_Ek_v04}}{\includegraphics[width=.49\linewidth]{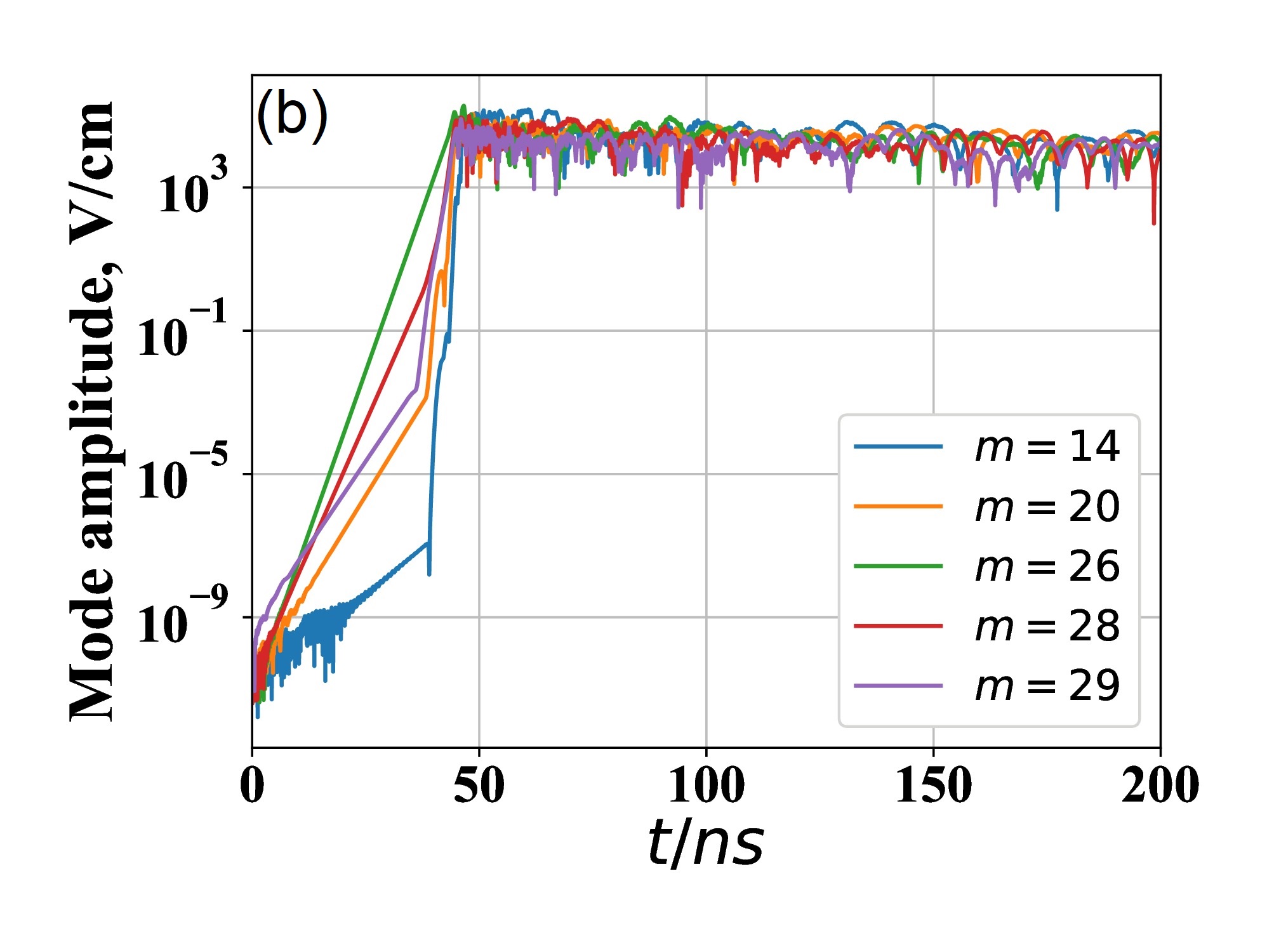}}
\caption{a) The electric field energy ($\mathcal{E}$) versus time. b) The evolution of the amplitude of the individual electric field modes. The case of $v_0=4v_{te}$.}
\end{figure}

\begin{figure}[htbp]
\centering
\captionsetup[subfigure]{labelformat=empty}
\subcaptionbox{\label{t_efE_v010}}{\includegraphics[width=.49\linewidth]{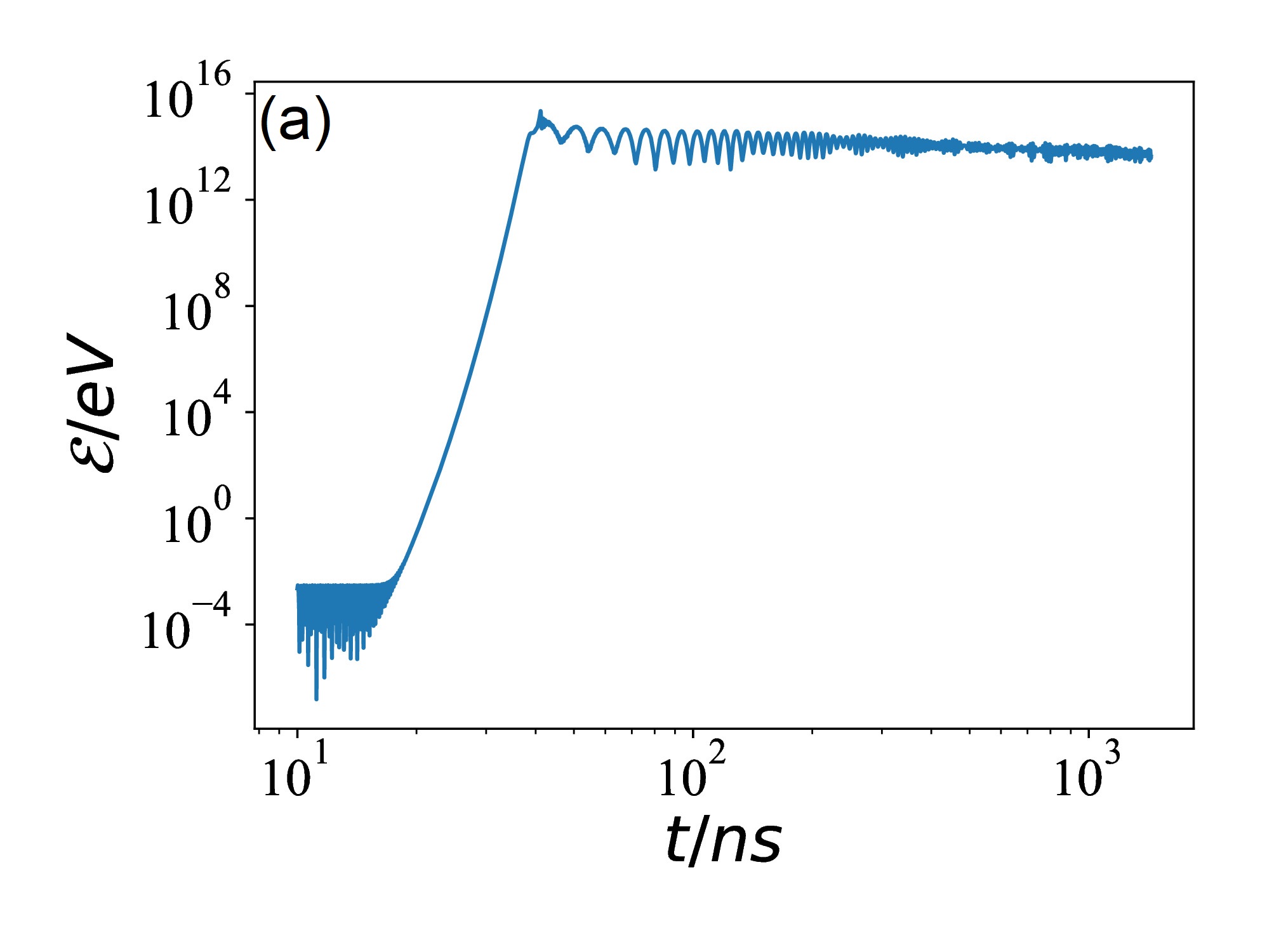}}
\subcaptionbox{\label{t_Ek_v010}}{\includegraphics[width=.49\linewidth]{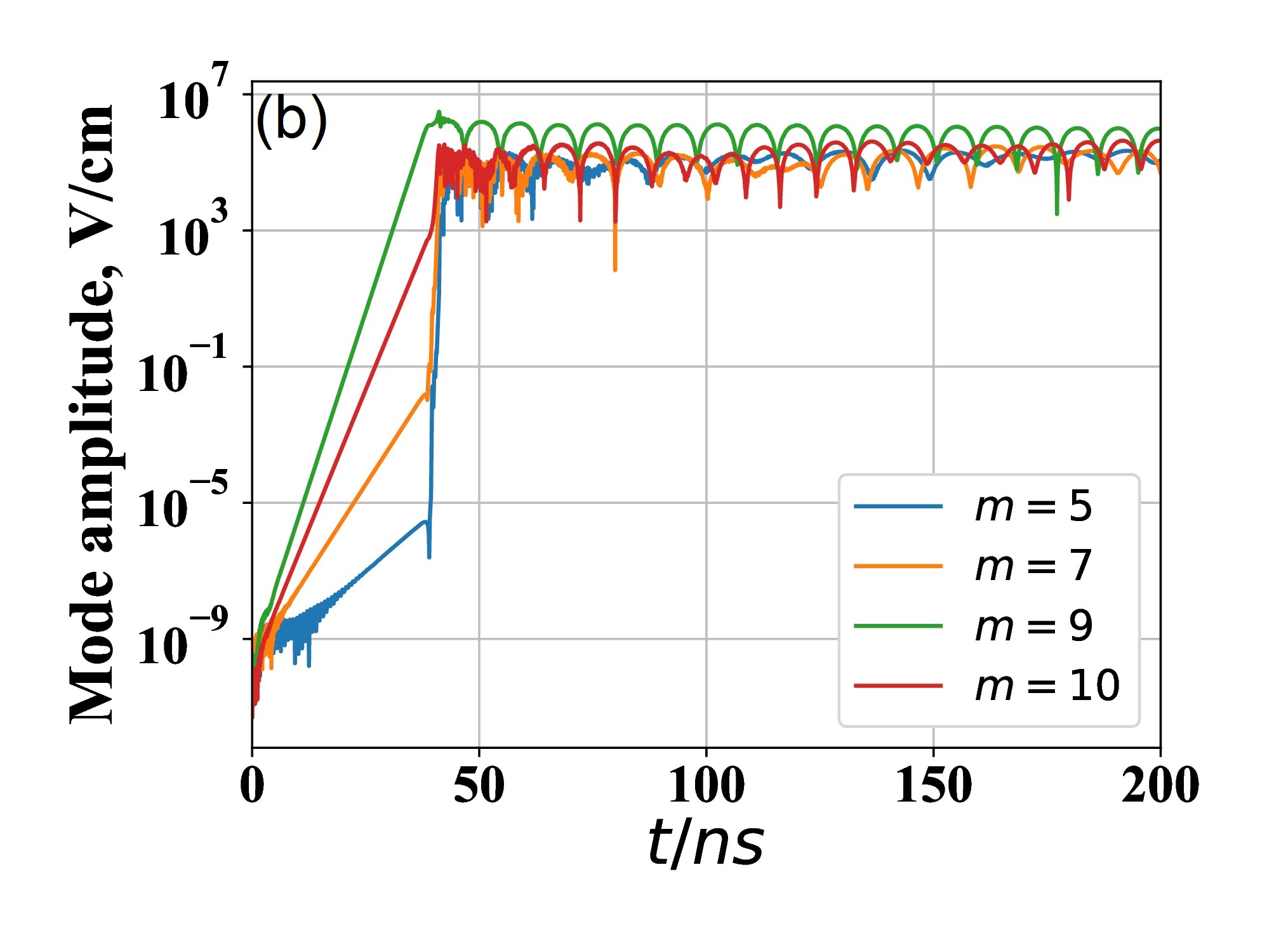}}
\caption{a) The electric field energy ($\mathcal{E}$) versus time. b) The evolution of the amplitude of the individual electric field modes. The case of $v_0=10v_{te}$.}
\end{figure}

\begin{figure}[htbp]
\centering
\captionsetup[subfigure]{labelformat=empty}
\subcaptionbox{\label{total_fe}}{\includegraphics[width=.8\linewidth]{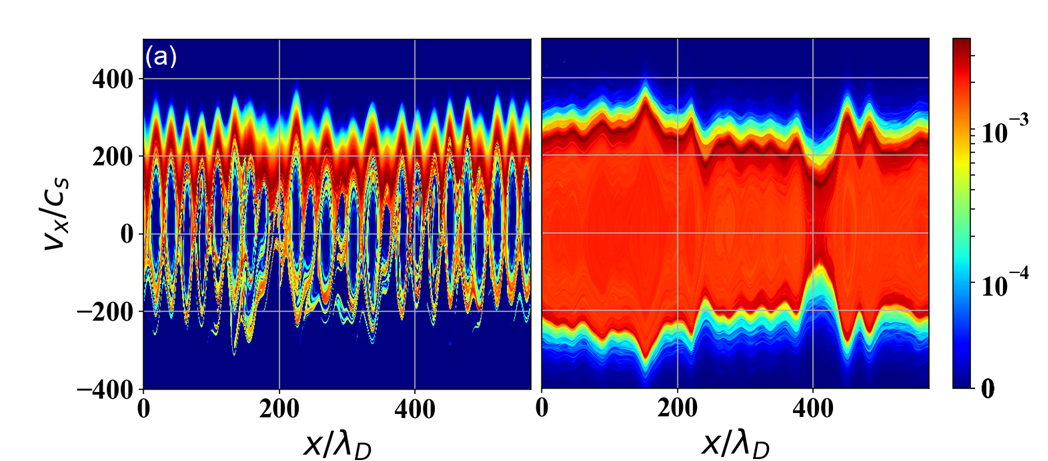}}
\subcaptionbox{\label{x_phi}}{\includegraphics[width=.75\linewidth]{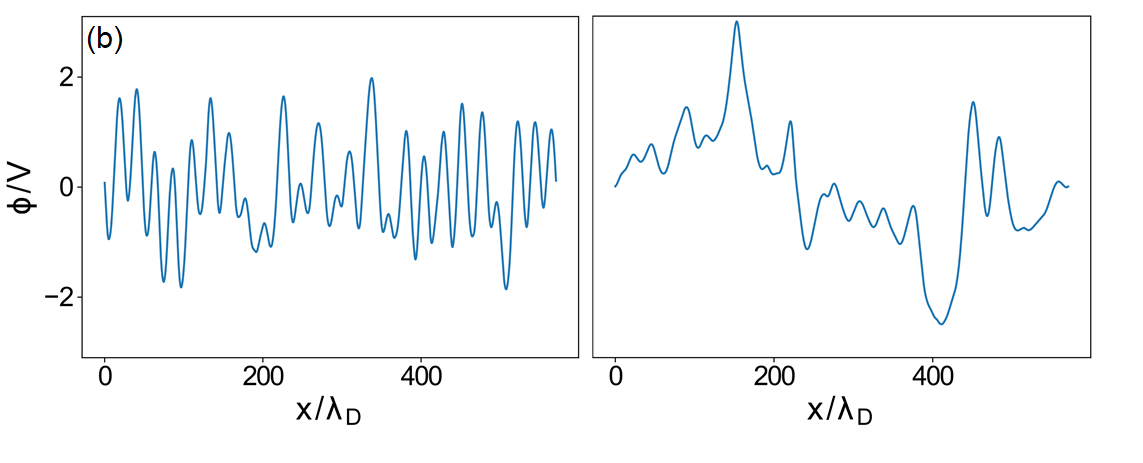}}
\subcaptionbox{\label{total_f}}{\includegraphics[width=.8\linewidth]{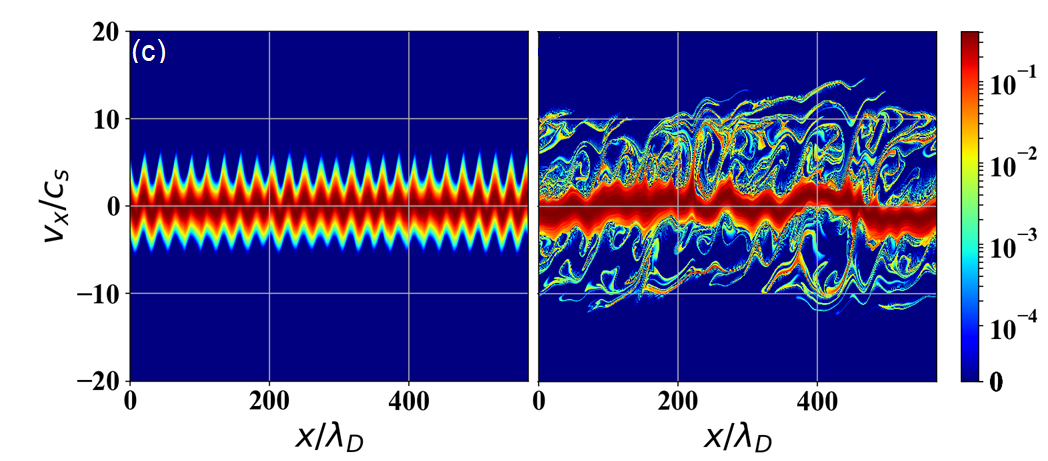}}
\caption{(a) The electron distribution function in $n_0/c_s$, (b) Electrostatic potential, (c) The ion distribution function in $n_0/c_s$. In each row, the figure on the left-hand side is at $t= 47.39$ ns ($19.8\;\omega_{pi}^{-1}$), and the figure on the right-hand side is at $t= 163.22$ ns ($68.2\;\omega_{pi}^{-1}$). These figures correspond to the case of $v_0=4v_{te}$. }
\end{figure}

After a few nanoseconds of simulation,
the electric field energy starts to grow linearly and continues until it reaches a peak. This peak is followed by a slow decay due to the energy transfer from the waves to the plasma and particle heating (\Cref{t_efE_v04,t_efE_v010}). The nonlinear regime is characterized by the appearance of trapping holes in the ion and electron distribution functions. The ion holes correspond to negative electrostatic potential, whereas the electron holes correspond to positive electrostatic potential\cite{hutchinson2017electron}. The electron holes appear early in the nonlinear regime and in the bulk region of the distribution function (\Cref{total_fe}, left). As we see in the left figure in \Cref{total_fe}, the number of these holes in the early nonlinear regime is close to the mode number of the most unstable mode (26, in the case of $v_0=4v_{te}$). The left figure in \Cref{x_phi}  shows the potential profile at the same time as the left figure in \Cref{total_fe}. The large-amplitude (rogue) waves about $1.5$ V to $2$ V can be seen in this figure. This amplitude is of the same order of magnitude as the initial drift energy of electrons ($1/2m_ev_0^2 \approx 1.6$ eV).  Early electron holes then merge together and form larger holes (\Cref{total_fe}, right) in a process discussed by a number of previous works\cite{ghizzo1988stability}. This process of merging leads to the appearance of higher wavelengths in the potential profile (\Cref{x_phi}, right). In addition, the ion holes appear in the tail of the distribution function later in the nonlinear regime (\Cref{total_f}, right). Backward-propagating waves appear in the early nonlinear regime after the wave amplitude is large enough to significantly reflect the electrons backward and extend the plateau to the negative regions of electron VDF. Backward and forward waves can be seen clearly in \Cref{total_ef_v04,total_ef_v010}.

\begin{figure}[htbp]
\centering
\includegraphics[width=.49\linewidth]{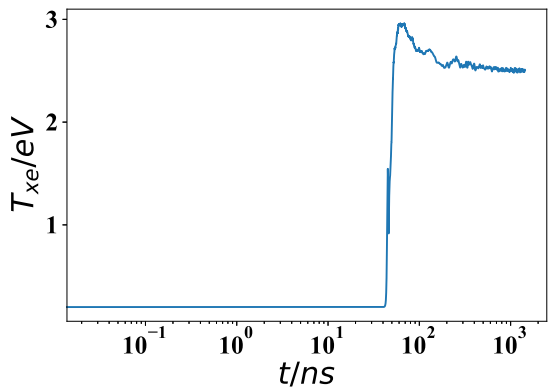}
\caption{The evolution of electron temperature for the case $v_0=4v_{te}$}
\label{t_avg_Txe}
\end{figure}

\begin{figure}[htbp]
\centering
\includegraphics[width=0.46\linewidth]{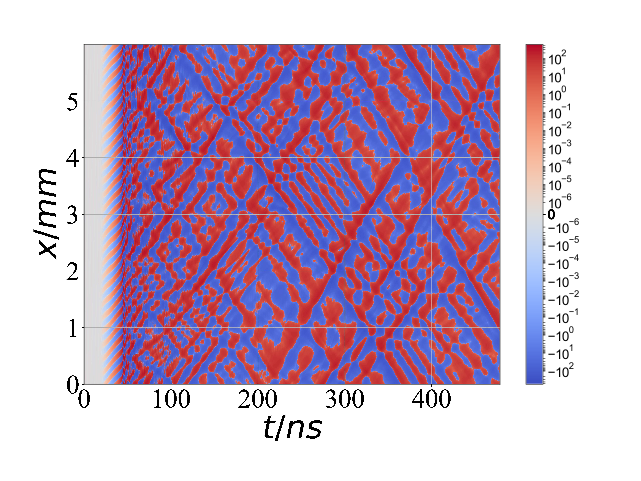}
\caption{The electric field (V/cm) as a function of time and position for the case of $v_0=4v_{te}$. After the backward-propagating waves appear at around $t=45$ ns, the coexistence of backward and forward waves forms a grid pattern in the electric field profile.}
\label{total_ef_v04}
\end{figure}

\begin{figure}[htbp]
\centering
\includegraphics[width=0.46\linewidth]{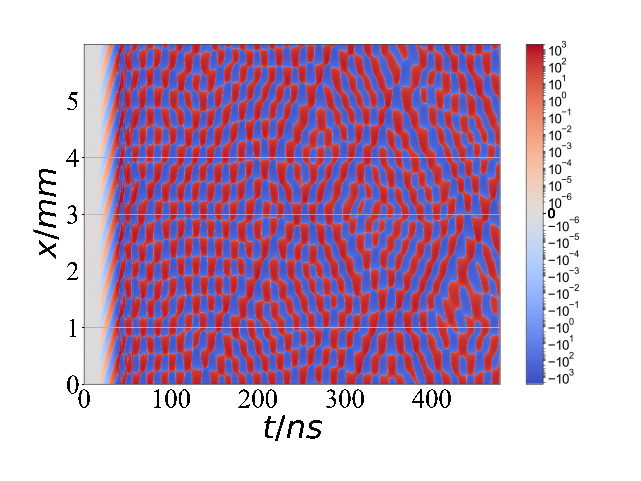}
\caption{The electric field (V/cm) as a function of time and position for the case $v_0=10\,v_{te}$. After the backward-propagating waves appear at around $t=30$ ns, the coexistence of backward and forward waves forms a grid pattern in the electric field profile.}
\label{total_ef_v010}
\end{figure}

The electron  scattering from large-amplitude fluctuations of the potential results in a dramatic  increase of the electron temperature. \Cref{t_avg_Txe} shows the evolution of the (spatially averaged) temperature  
\begin{equation}
   T_e= \frac{1}{n_0L} \int m_e(v_x-V_{xe})^2f_e(x,v_x,t)\, \dd x\, \dd v_x.
\end{equation}
The fluid velocity $V_{xe}$ is the first moment of distribution function; i.e.,~$V_{xe}=\int v_xf_e \, \dd v_x/\int f_e\, \dd v_x$.
The electron energy starts to increase in early nonlinear stage  (after about 47 ns), at the same time when the rogue waves appear in the potential and the backward waves are generated. The electron temperature then saturates to about $2.4$ eV. We note that this value is of the same order of magnitude as the amplitudes of the sharp peaks in \Cref{x_phi}.

The rest of the paper is dedicated to the characteristics of the spectra (backward and forward waves) in the nonlinear stage obtained in simulations and from analytical calculations. Before that, however, it is interesting to note another nonlinear phenomenon that can be seen in our results. In all of our simulations, we see a significant group of electrons that have been accelerated ahead of the initial drift velocity, as seen in the electron VDF (see for example the small bump and energetic tail in the right-hand side of VDF, in \Cref{avg_fe_v04,avg_fe_v10}). This acceleration is likely due to the electron trapping and de-trapping in the large-amplitude (forward) waves. Similar self-acceleration of the beam in the two-stream instabilities has  been studied  earlier in Ref.~\cite{HaraPoP2018,Xu2020self}. 

\section{Backward waves as marginally stable eigen-modes of the nonlinearly modified velocity distribution function}\label{analysis}

\begin{figure}[htbp]
\centering
\captionsetup[subfigure]{labelformat=empty}
\subcaptionbox{\label{fe0_v04}}{\includegraphics[width=0.46\linewidth]{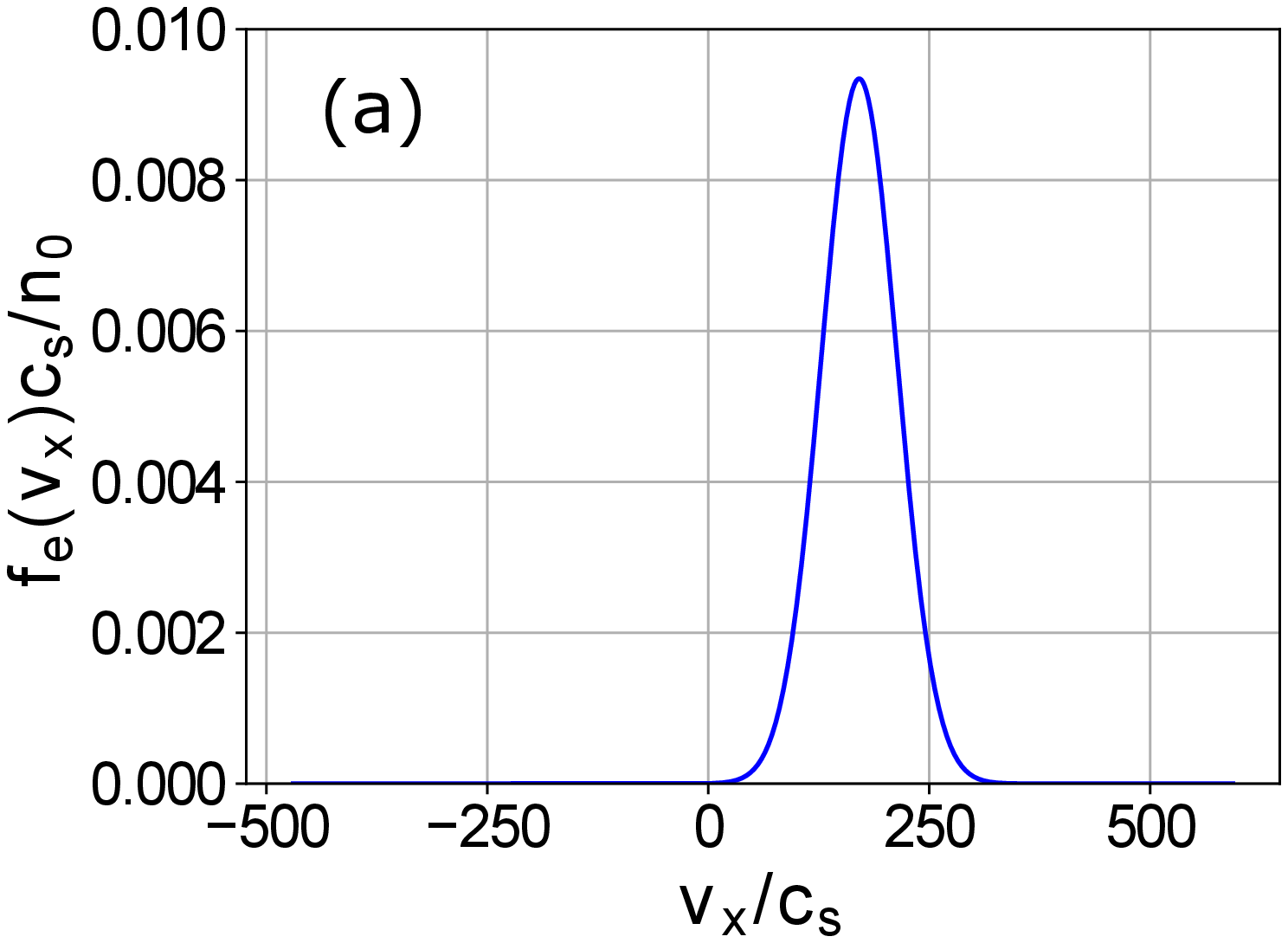}}\vspace{-2em} \par\medskip
\subcaptionbox{\label{avg_fe47.386to83.285}}{ \includegraphics[width=0.46\linewidth]{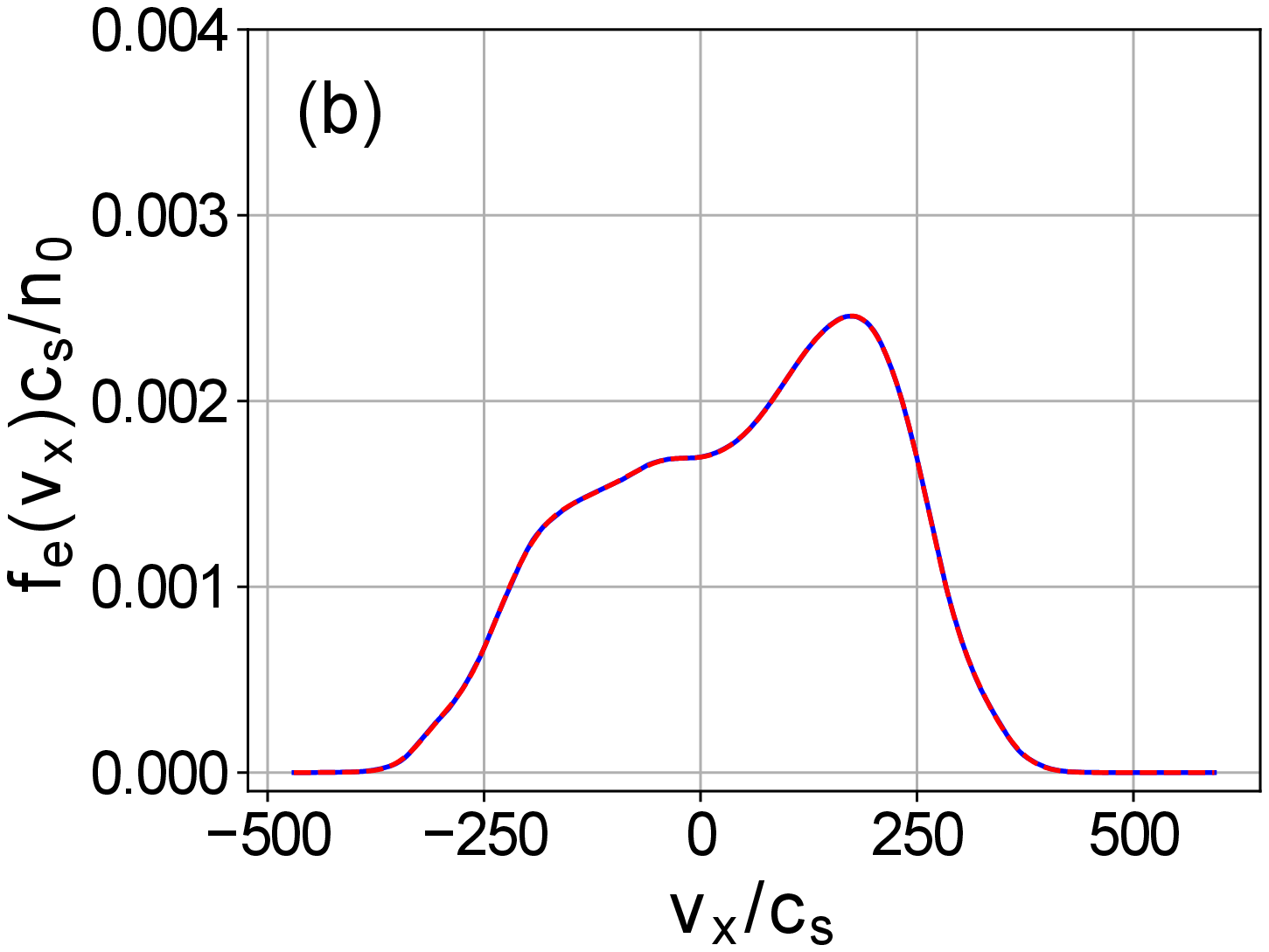}}\vspace{-2em}
\subcaptionbox{\label{avg_fe83.285to123.013}}{\includegraphics[width=0.46\linewidth]{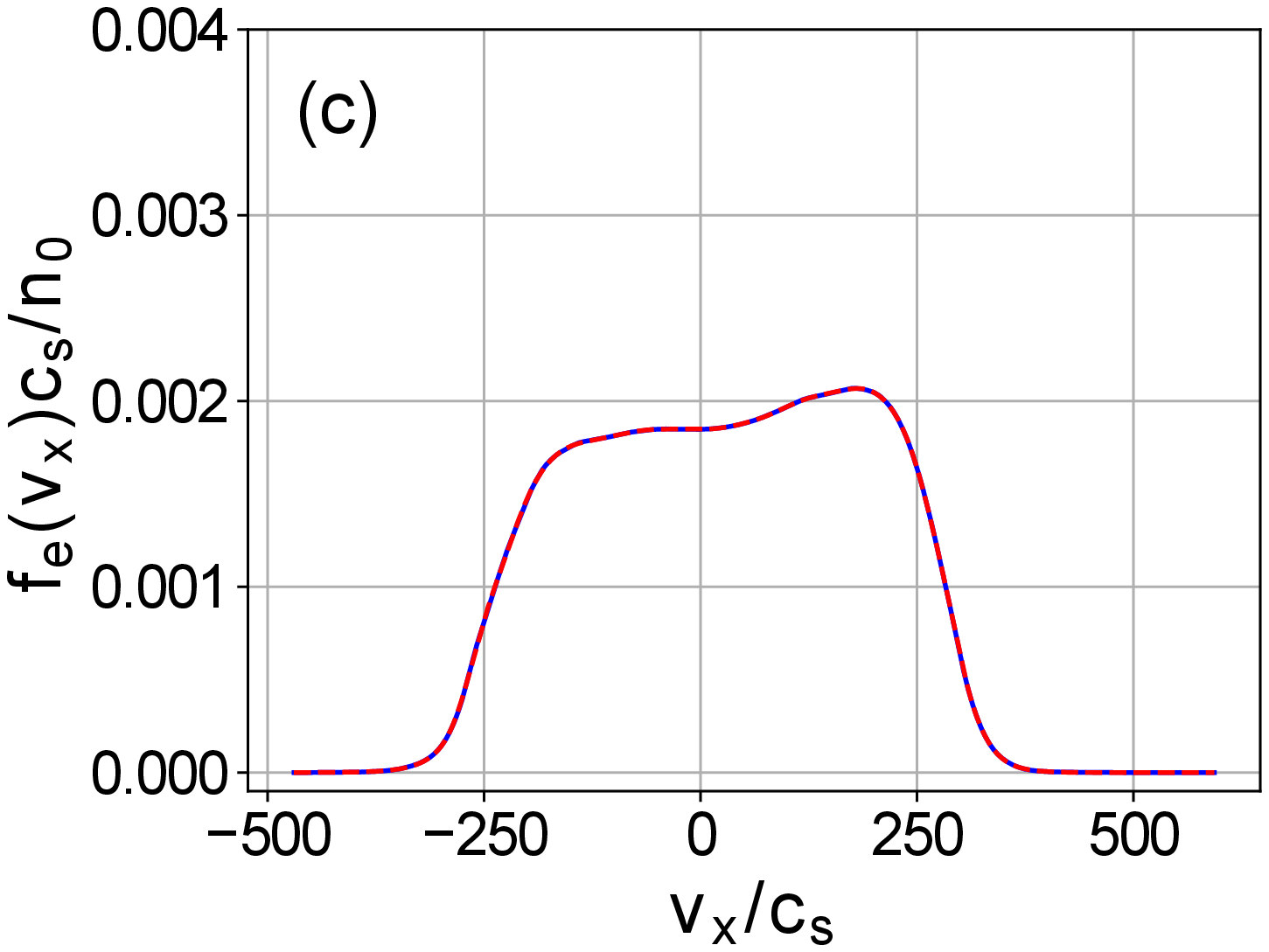}} \vspace{-2em}
\subcaptionbox{\label{avg_fe123.013to163.220}}{\includegraphics[width=0.46\linewidth]{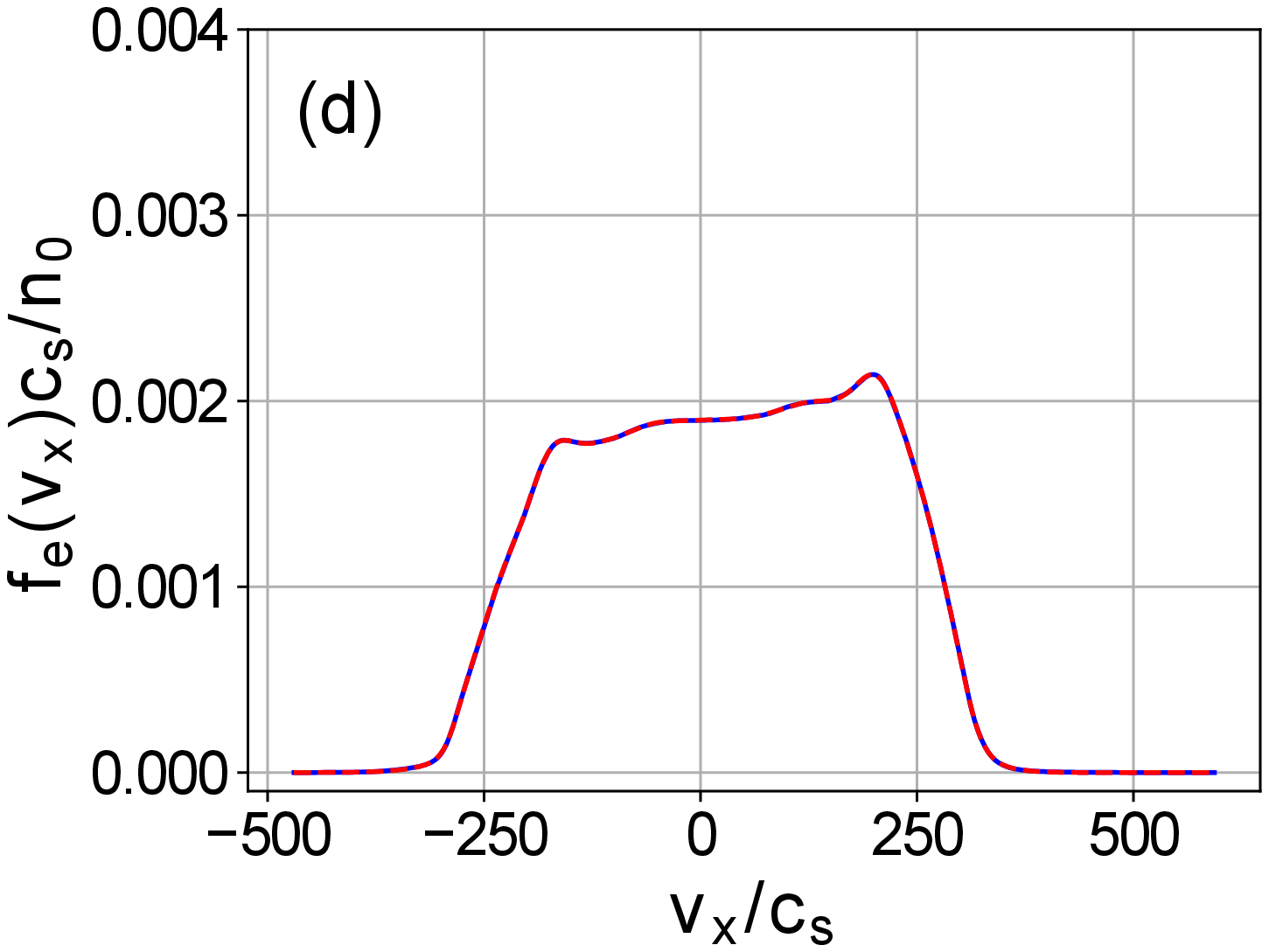}}
\subcaptionbox{\label{avg_fe163.220to239.325}}{\includegraphics[width=0.46\linewidth]{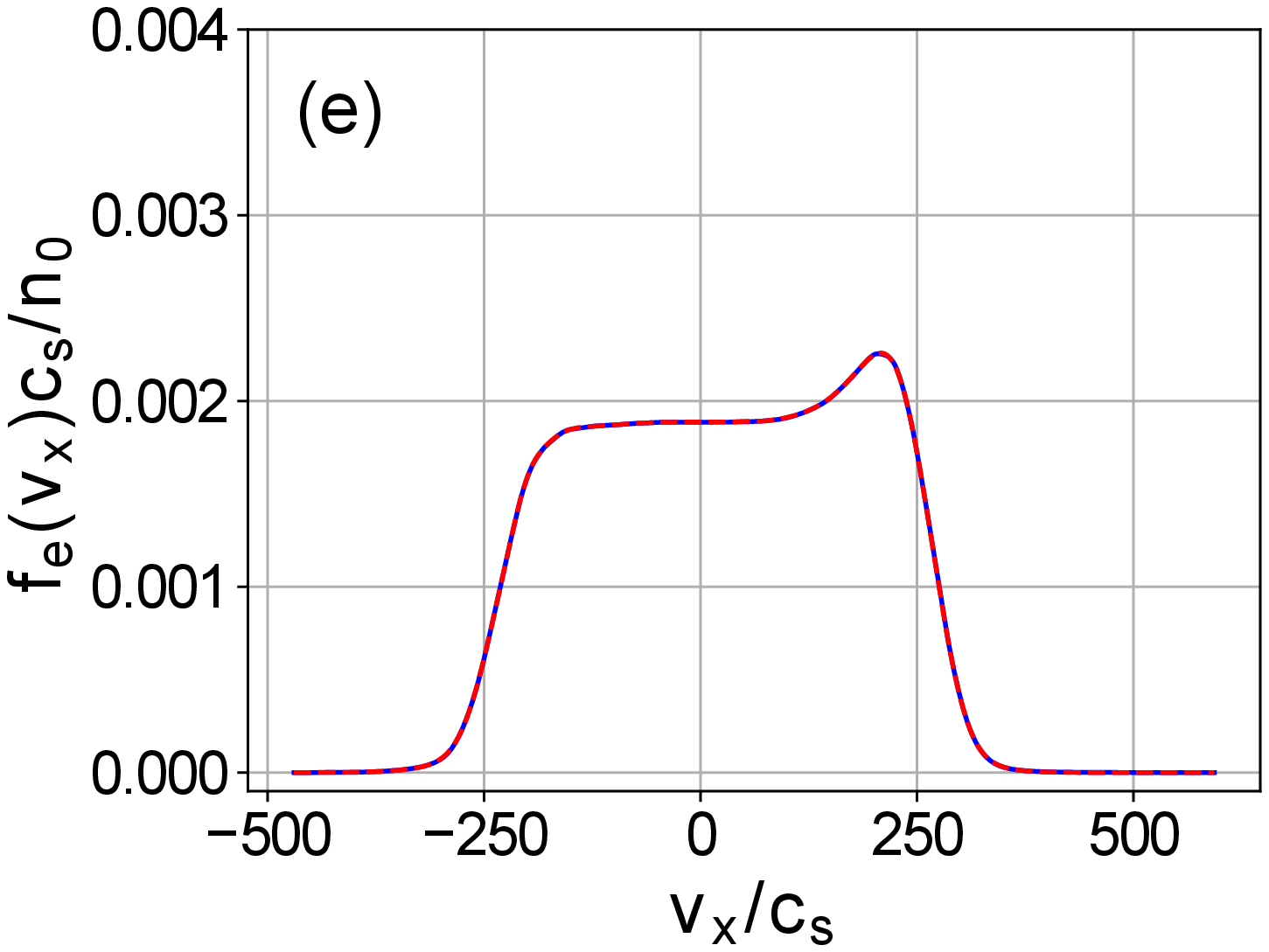}}
\caption{Evolution of the electron VDF for  $v_0=4\,v_{te}$. (a) the initial VDF at $t=0$  (b) VDF from nonlinear  simulations (blue line)  averaged  over  $19.8\,\omega_{pi}^{-1}$ to $34.8\,\omega_{pi}^{-1}$ (c) $34.8\,\omega_{pi}^{-1}$ to $51.4\,\omega_{pi}^{-1}$ (d) $51.4\,\omega_{pi}^{-1}$ to $68.2\,\omega_{pi}^{-1}$ (e) $68.2\,\omega_{pi}^{-1}$ to $100\,\omega_{pi}^{-1}$. The fit from \Cref{fittingVDF} is shown in red in (b),(c),(d), and (e).} 
\label{avg_fe_v04}
\end{figure}

As discussed above, the electron distribution in the nonlinear stage is strongly modified. To study the stability of such distributions, we represent the electron VDF obtained from the simulation by ten electron beams, each with a beam density of $\delta_j$, in the form
\begin{align}
    f_e(v_x)=\sum_{j=1}^{10}\frac{n_0\delta_j }{\sqrt{2\pi}v_{ej}}\exp(-\frac{(v_x-v_{0j})^2}{2v_{ej}^2}).
    \label{fittingVDF}
\end{align}
The $\delta_j$ are normalized by the condition $\sum_{j=1}^{10}\delta_j$ = 1. \Cref{avg_fe_v04} shows the evolution of the spatially averaged electron VDF in the case $v_0=4v_{te}$.  It can be seen that in each time interval, the VDFs of \Cref{fittingVDF} can be closely fit to the simulated VDFs. The fit is done using the SciPy \textit{curve\_fit} function, which uses the least-squares method to non-linearly fit the ten-Maxwellian VDF~\Cref{fittingVDF} to the simulated VDF. The positivity of the beam densities ($\delta_j$) and beam thermal velocities ($v_{ej}$) was enforced in the fit. We note that some of the beam velocities $v_{0j}$ in the fit have a negative value, which is important for the fit to be well extended to the negative velocities. In general, the more Maxwellian functions we use for the fit, the more accurate it is. However,  using more than ten Maxwellians does not  significantly change the value of the standard error as returned by the fitness function. We have also checked that that lower error does not have much impact on the theoretical spectrum of the eigen-modes. Each fit results in a set of 29 independent parameters $\delta_j,v_{0j},v_{ej}$. Using these parameters, we solve the corresponding dispersion equation
\begin{gather}
    1-\frac{\omega_{pi}^2}{2k^2v_i^2}Z^{'}\qty(\frac{\omega}{\sqrt{2}\abs{k}v_{i}})-\sum_{j=1}^{10}\frac{\delta_j\omega_{pe}^2}{2k^2v_{ej}^2}Z^{'}\qty(\frac{\omega-kv_{0j}}{\sqrt{2}\abs{k}v_{ej}})
    =0
    \label{disper_v04}
\end{gather}
for the theoretical spectrum and growth rates of the waves.  For the case $v_0=4 v_{te}$, the ion distribution function does not change much, so we use the initial value of the ion temperature ($v_{ti}$) for $v_i$. For  $v_0=10 v_{te}$, we observe a slight modification of the ion distribution; it is described below. 

\subsection{Linear eigen-mode spectra for  $v_0=4v_{te}$}

\Cref{Ekw_47.386to83.285,Ekw_83.285to123.013,Ekw_123.013to163.220,Ekw_163.220to239.325} show the spectrum of nonlinear waves in four subsequent stages of nonlinear simulations. In each figure, we also show the eigen-modes obtained  from the solution of \Cref{disper_v04}. In general, the modes in two different regions can be seen in these figures: the arc-shaped high-frequency modes with $\omega \sim O(\omega_{pe})$ (Langmuir-like modes) and the low-frequency modes with $\omega\sim O(\omega_{pi})$ (ion-sound-like). We note that, for the FFT in time, we have used the Hanning window to reduce the amount of spectral leakage\cite{smith1974fast}. Despite this, some faint modes can be seen in between.  The time intervals were taken long enough so that the FFT has a relatively high resolution and can clearly show the low-frequency modes. As we see, the amplitudes of the low-frequency modes are several orders of magnitude larger than those of the high-frequency modes, and therefore, they are dominant  in  the nonlinear regime. The phase velocity of the low-frequency modes is close to the velocity of the ion holes in the ion distribution function (\Cref{total_f}, right). 
\begin{figure*}[htbp]
\centering
\captionsetup[subfigure]{labelformat=empty}
\subcaptionbox{\label{Ekw_47.386to83.285}}{\includegraphics[width=1\linewidth]{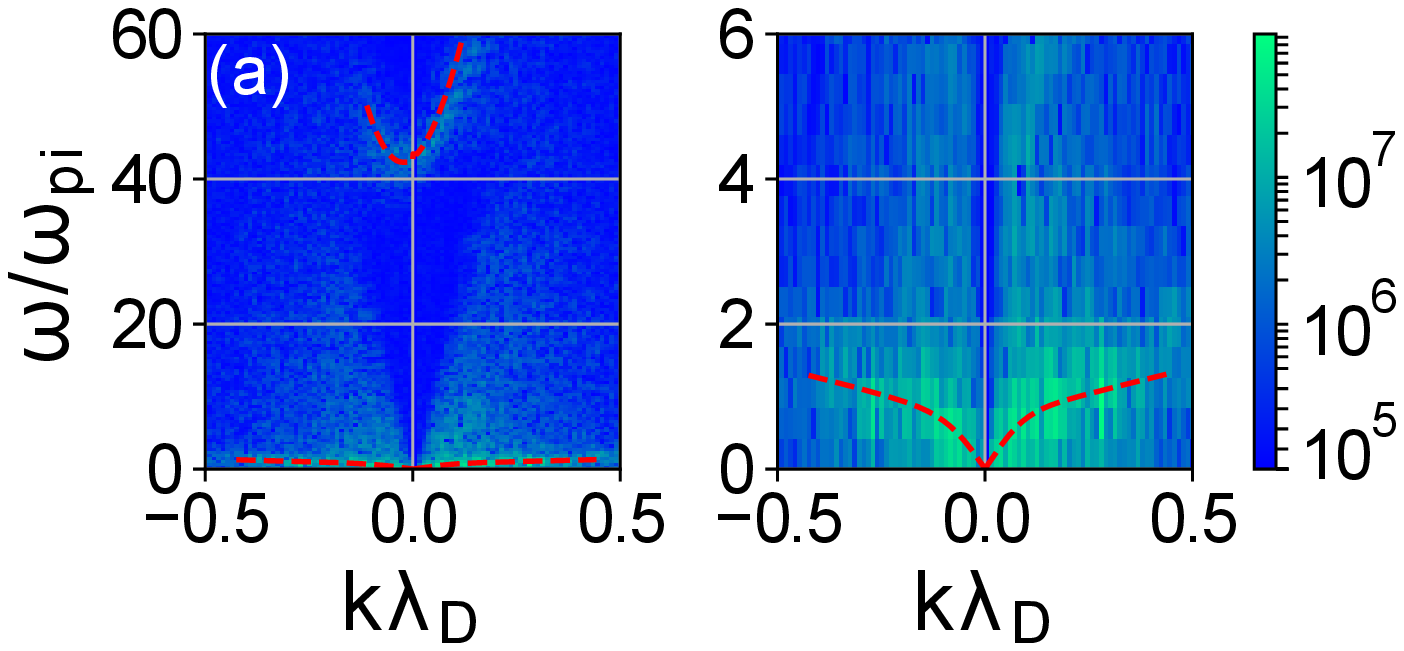}}\vspace{-2em}
\subcaptionbox{\label{Ekw_83.285to123.013}}{\includegraphics[width=1\linewidth]{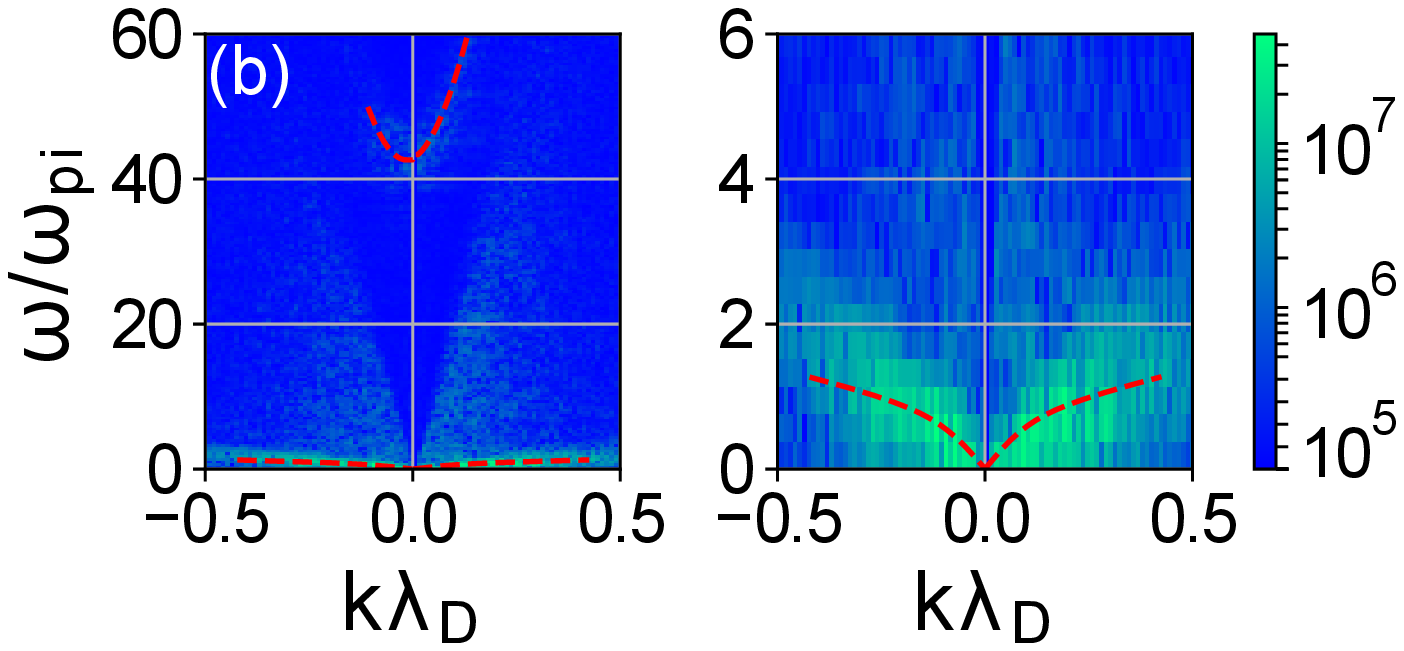}}\vspace{-2em}
\subcaptionbox{\label{Ekw_123.013to163.220}}{\includegraphics[width=1\linewidth]{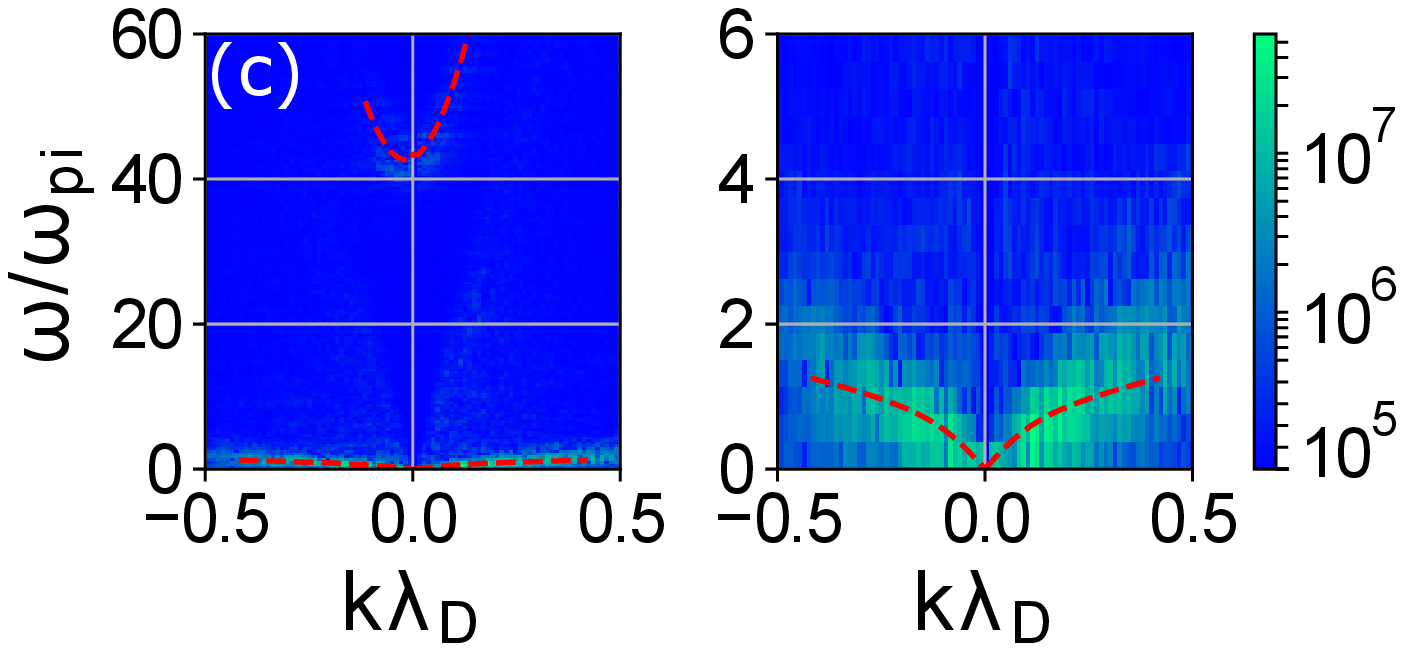}}\vspace{-3em}
\end{figure*}
\begin{figure*}[htbp]
\captionsetup[subfigure]{labelformat=empty}
\ContinuedFloat
\subcaptionbox{\label{Ekw_163.220to239.325}}{\includegraphics[width=1\linewidth]{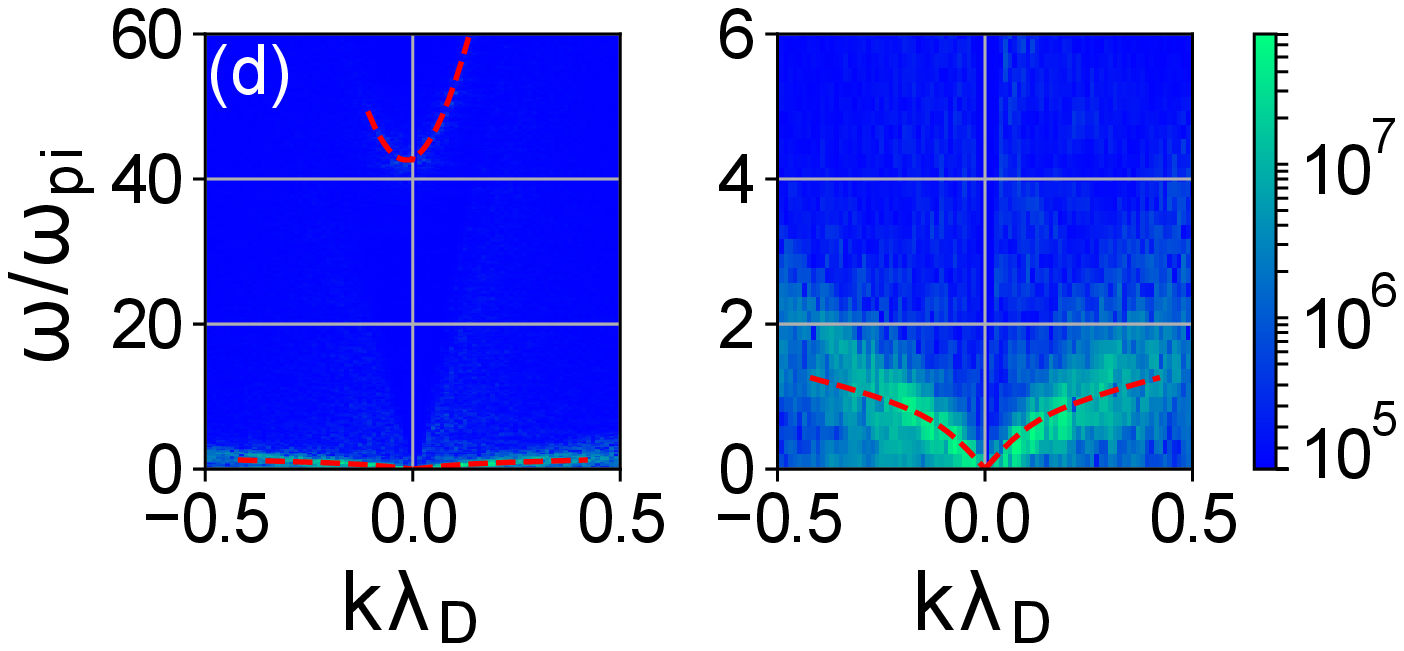}}
\caption{The spectrum of nonlinear waves in the case of $v_0=4v_{te}$, for (a) $19.8\,\omega_{pi}^{-1}$ to $34.8\,\omega_{pi}^{-1}$, (b) $34.8\,\omega_{pi}^{-1}$ to $51.4\,\omega_{pi}^{-1}$, (c) $51.4\,\omega_{pi}^{-1}$ to $68.2\,\omega_{pi}^{-1}$, (d) $68.2\,\omega_{pi}^{-1}$ to $100\,\omega_{pi}^{-1}$. In each case, a zoom into the low-frequency region is shown on the right of the the full spectrum. The red lines show the modes found by solving \Cref{disper_v04}.}
\label{Ekw_v04}
\end{figure*}

\begin{figure*}[htbp]
\centering
\captionsetup[subfigure]{labelformat=empty}
\subcaptionbox{\label{disp_47.386to83.285}}{\includegraphics[width=0.46\linewidth]{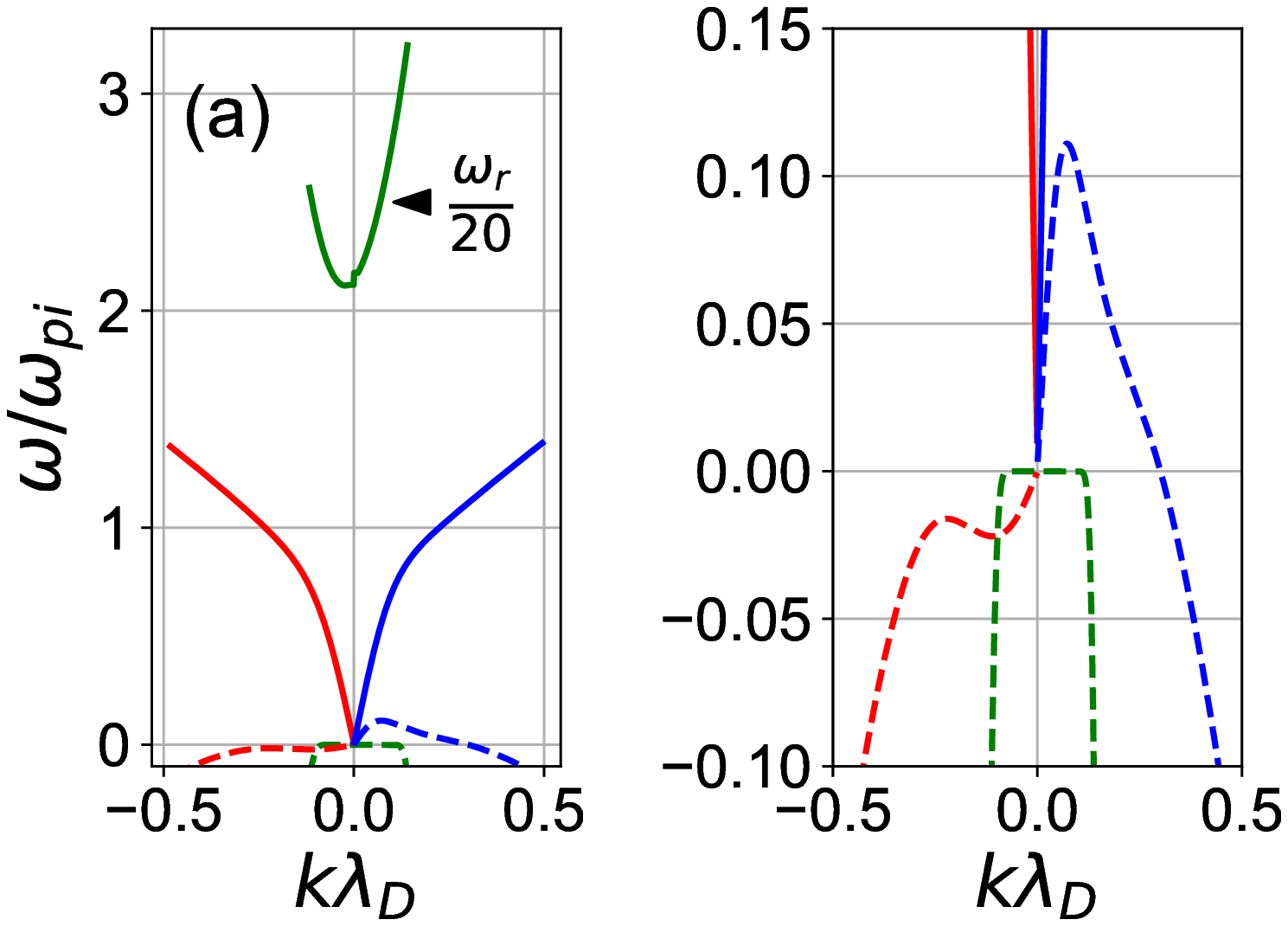}}\vspace{-1em}
\subcaptionbox{\label{disp_83.285to123.013}}{\includegraphics[width=0.46\linewidth]{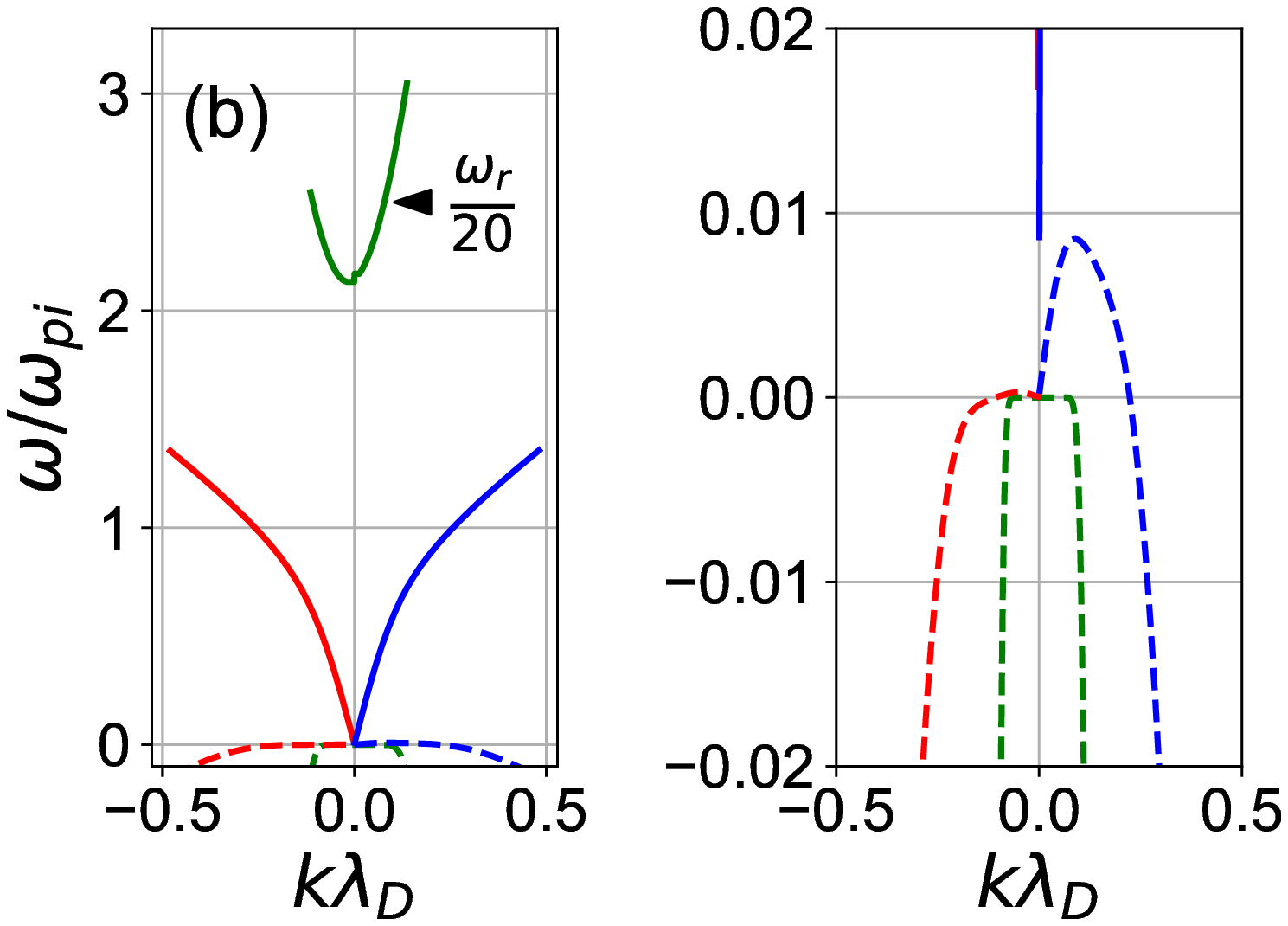}}\vspace{-1em}
\subcaptionbox{\label{disp_123.013to163.219}}{\includegraphics[width=0.46\linewidth]{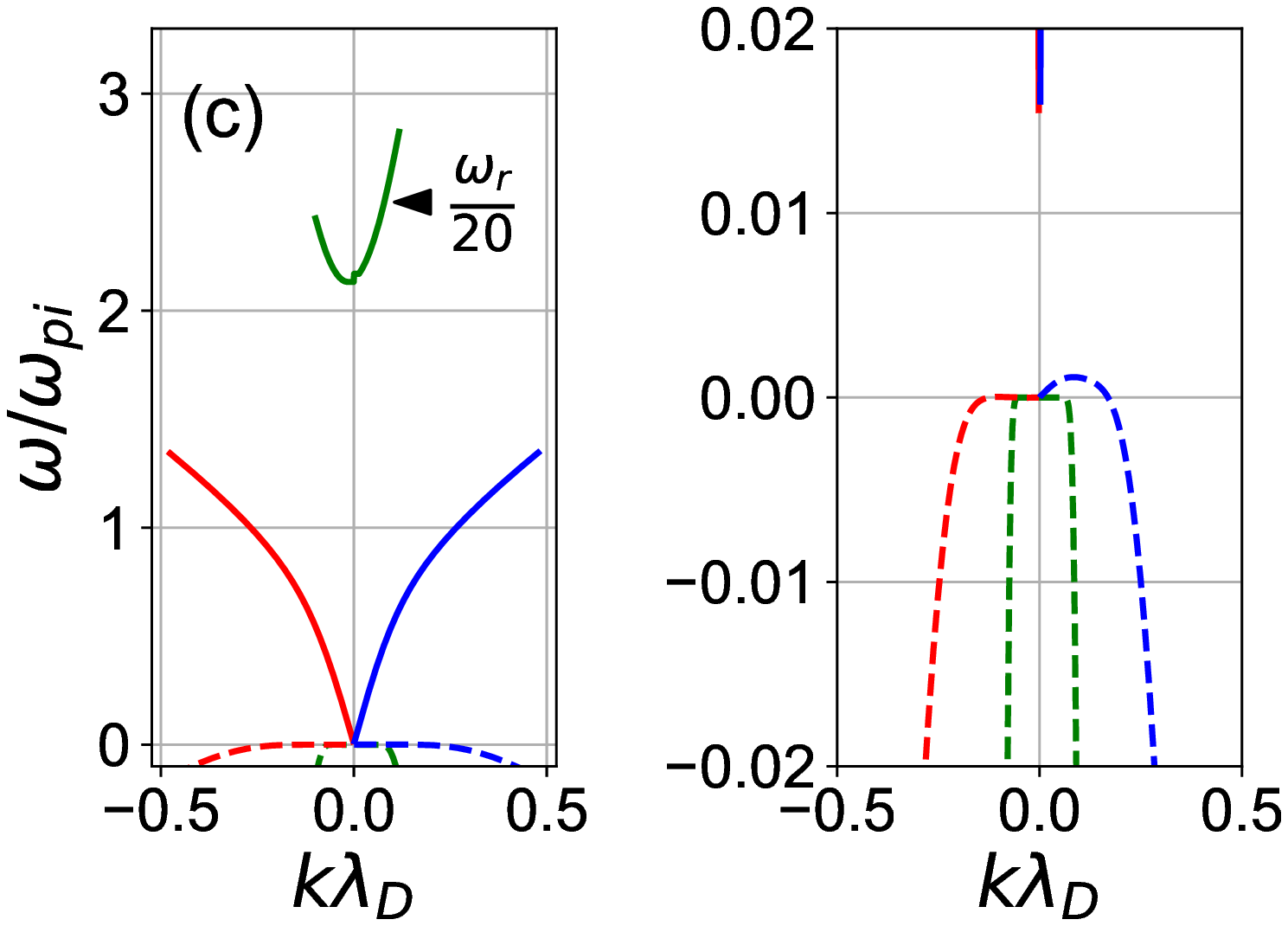}}\vspace{-1em}
\subcaptionbox{\label{disp_163.219to239.325}}{\includegraphics[width=0.46\linewidth]{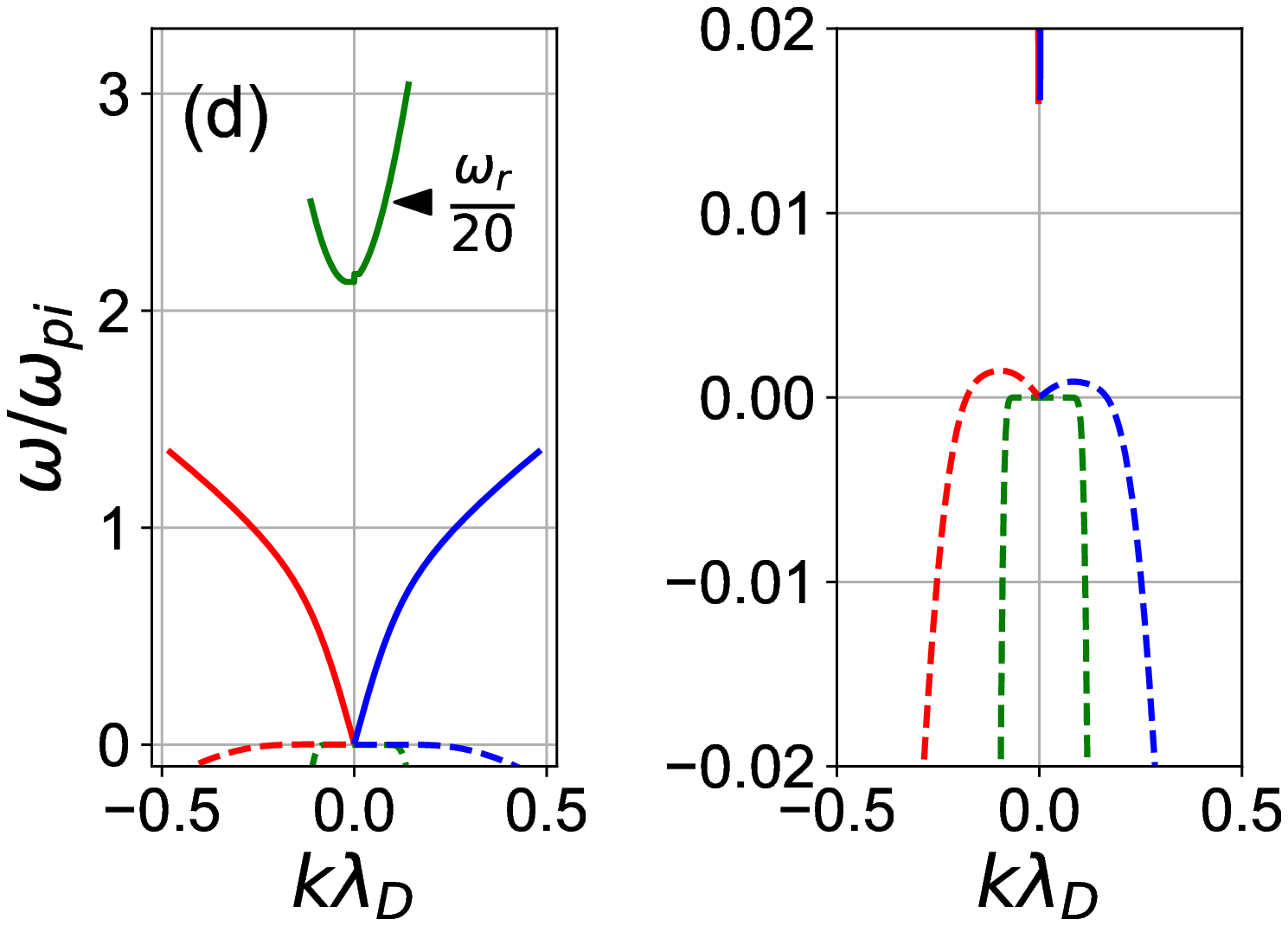}}\vspace{-1em}
\caption{The theoretical frequencies (solid lines) and the growth rates (dashed lines) for the case of $v_0=4v_{te}$ shown in four stages. (a) $19.8\,\omega_{pi}^{-1}$ to $34.8\,\omega_{pi}^{-1}$, (b) $34.8\,\omega_{pi}^{-1}$ to $51.4\,\omega_{pi}^{-1}$, (c) $51.4\,\omega_{pi}^{-1}$ to $68.2\,\omega_{pi}^{-1}$, and (d) $68.2\,\omega_{pi}^{-1}$ to $100\,\omega_{pi}^{-1}$. These results are found by solving \Cref{disper_v04}. The modes with growth rates less than $-0.1 \omega_{pi}$ are omitted. The real and imaginary parts of a complex root are shown with the same color.}
\label{disp_v04}
\end{figure*}

\Cref{disp_47.386to83.285} to \Cref{disp_163.219to239.325} show the theoretical mode frequencies $\omega$ and corresponding growth rates $\gamma$ for the distribution function from the corresponding time intervals of Fig. 9.  In these figures, we have omitted growth rates less than $-0.1\omega_{pi}$. 
We can see small positive or near-zero growth rates in the range of $\abs{k\lambda_D}\lesssim 0.4$, whereas the modes outside of this range are  strongly damped due to the ion Landau damping, which is significant for phase velocities around $O(v_{ti})$. These spectra of the weakly unstable or  marginally stable eigen-modes  in the range    $\abs{k\lambda_D}\lesssim 0.4$ show close resemblance to the eigen-mode spectra obtained in the nonlinear simulations  (\Cref{Ekw_47.386to83.285} to \Cref{Ekw_163.220to239.325}). 

In \Cref{disp_v04}, we observe the  asymmetry between positive and negative $k$ for the Langmuir-like (high-frequency) modes. This asymmetry is a result of asymmetric damping of the positive and negative $k$ modes, and it can also be seen in the simulation results (\Cref{Ekw_v04}). Our theoretical analysis of the eigen-mode spectra of the modified distribution function also shows these high-frequency modes with near-zero or  negative growth rates as seen in \Cref{disp_47.386to83.285} to \Cref{disp_163.219to239.325}. Similarly, in our simulations, these modes fade away further as seen in the last time window (\Cref{Ekw_163.220to239.325}).

\subsection{Linear eigen-mode spectra in the cold plasma limit of the Buneman instability ($v_0=10v_{te}$)}

\begin{figure}[htbp]
\centering
\includegraphics[width=.49\linewidth]{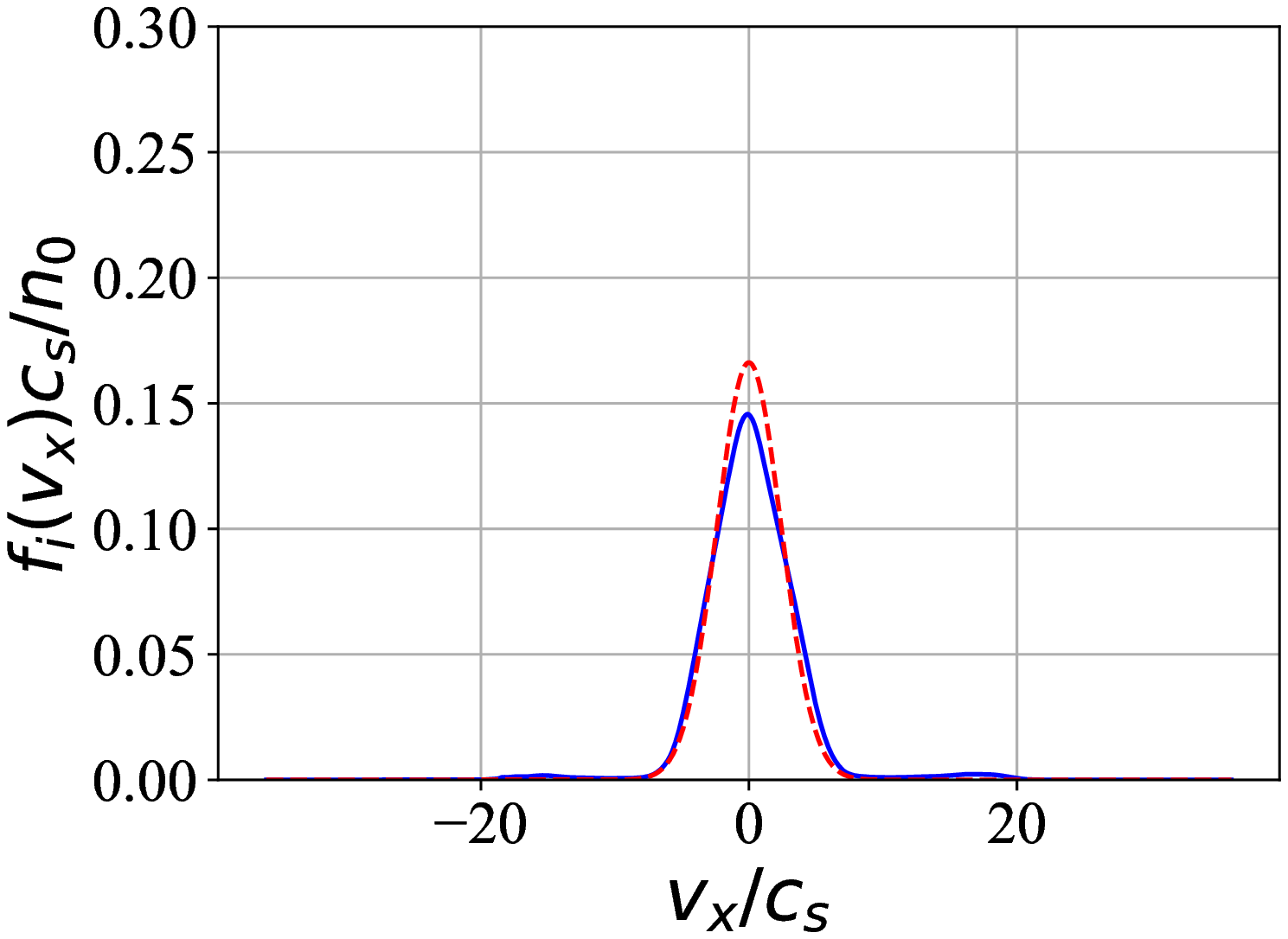}
\caption{Ion distribution function for the case $v_0=10\,v_{te}$, averaged in $31.4\,\omega_{pi}^{-1}$ to $46\,\omega_{pi}^{-1}$ (solid blue line) shown together with the fitted Maxwellian (dashed red line).}
\label{avg_f75.148to110.090}
\end{figure}

\begin{figure}[htbp]
\centering
\captionsetup[subfigure]{labelformat=empty}
\subcaptionbox{\label{fe0_v010}}{ \includegraphics[width=0.4\linewidth]{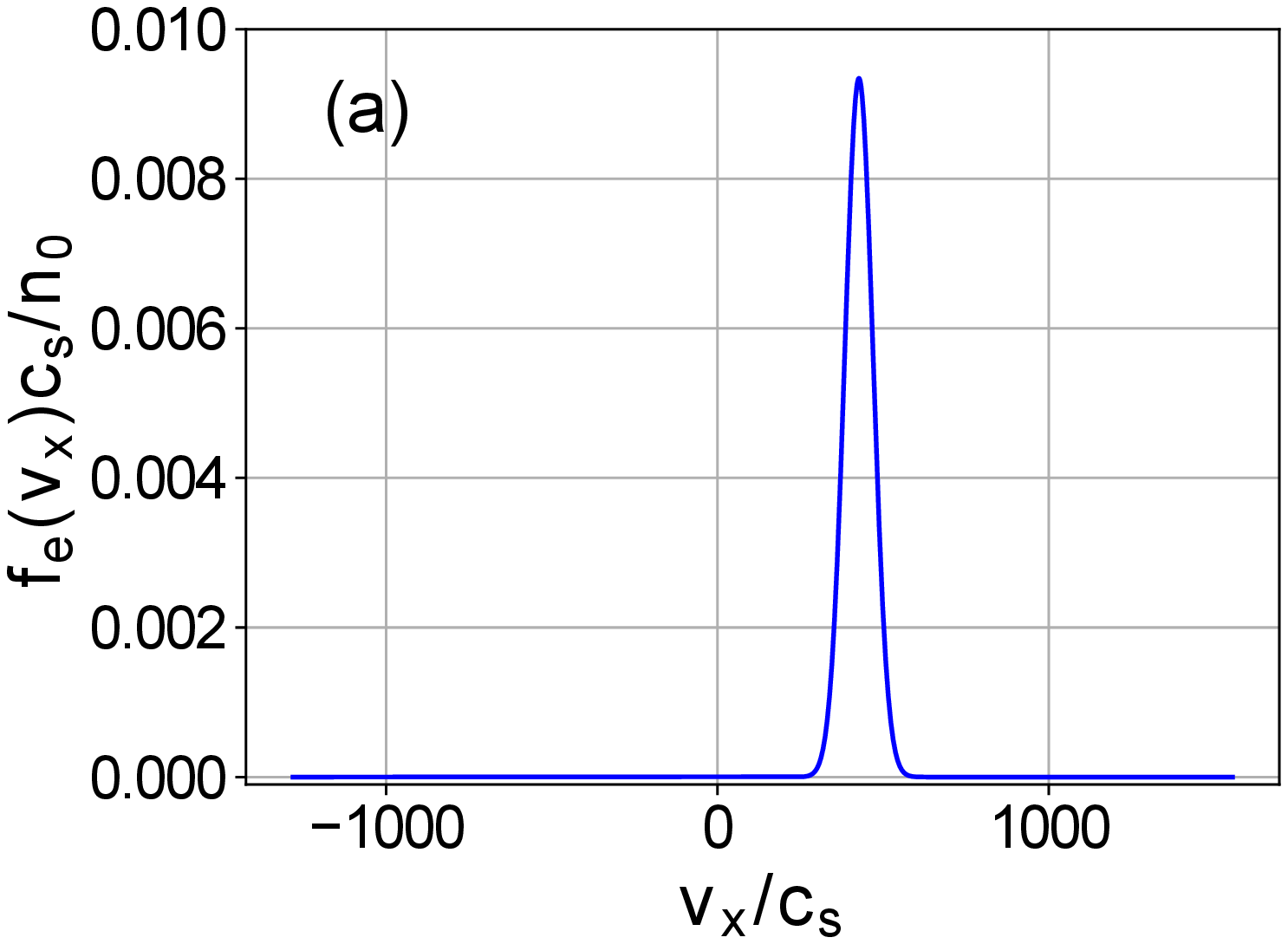}}\vspace{-1em}
\subcaptionbox{\label{avg_fe40.207to75.148}}{\includegraphics[width=0.4\linewidth]{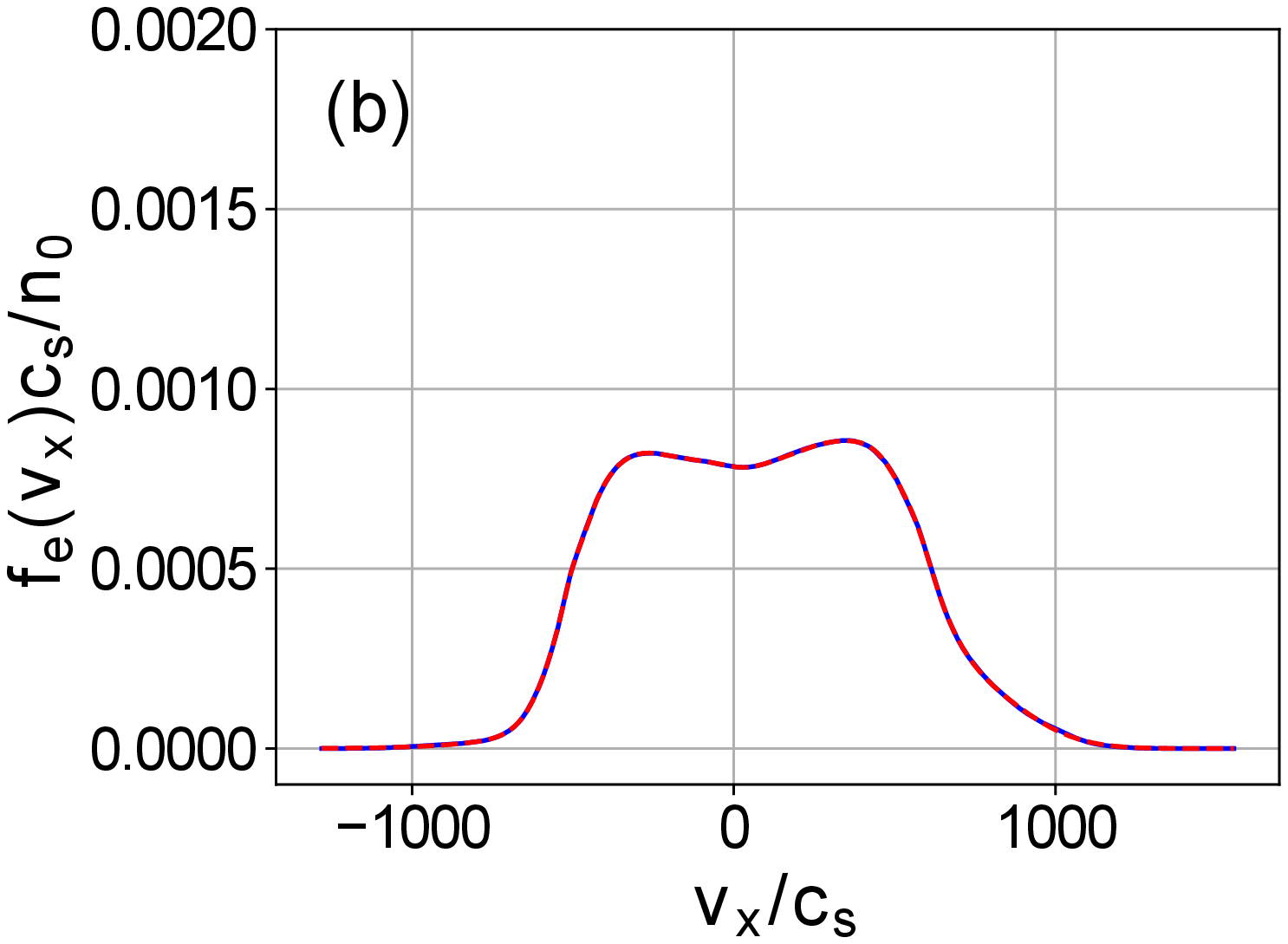}}\vspace{-1em}
\subcaptionbox{\label{avg_fe75.148to110.090}}{\includegraphics[width=0.4\linewidth]{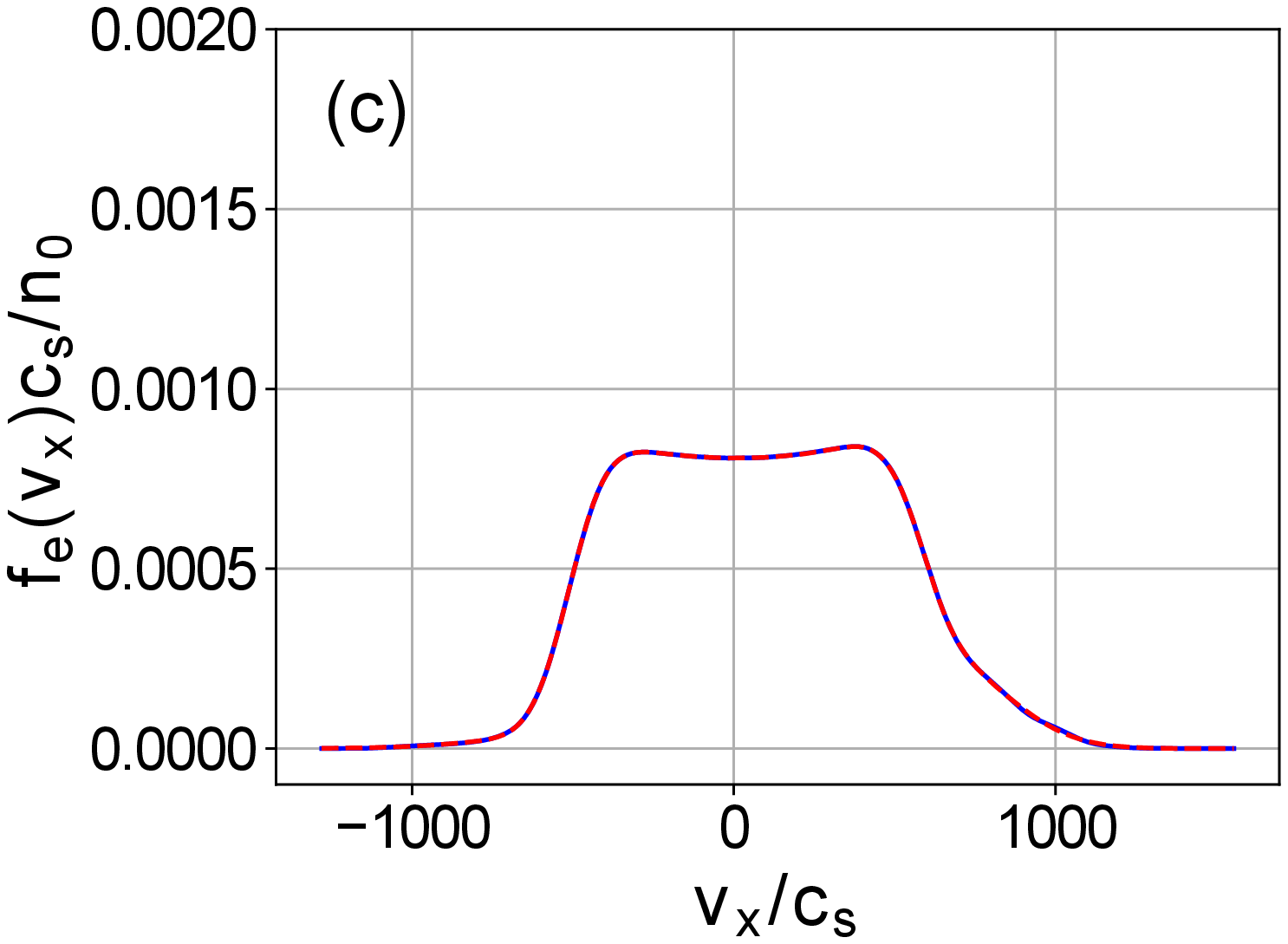}}\vspace{-3em}
\caption{Evolution of the electron VDF for  $v_0=10\,v_{te}$. (a) the initial VDF at $t=0$  (b) VDF from nonlinear  simulations (blue line)  averaged  over  $t=16.8\,\omega_{pi}^{-1}$ to $31.4\,\omega_{pi}^{-1}$ (c) $t=31.4\,\omega_{pi}^{-1}$ to $46\,\omega_{pi}^{-1}$. The fit from \Cref{fittingVDF} is shown in red in (b) and (c).}
\label{avg_fe_v10}
\end{figure}

Increasing the initial electron drift to $v_0=10v_{te}$ leads to a stronger instability, which saturates to a much higher value of electric field energy (\Cref{t_efE_v010}) and increasing  the electron and ion heating. In order to explain the nonlinear modes in this case, we need to consider the ion heating in the nonlinear regime. We therefore fit a Maxwellian function to the ion VDF in one of the time windows, and find $v_{ti}=2.4\,c_s$ for use in \Cref{disper_v04} (\Cref{avg_f75.148to110.090}). We note that it is important to have the most accurate fit in the region of the phase velocity of propagating waves. Therefore, in calculating the fit residual, we have given a special weight to the points around that region ($v_x\approx \pm 10\;c_s$ in \Cref{avg_f75.148to110.090}). The electron VDF is also fit in two time windows (\Cref{avg_fe40.207to75.148,avg_fe75.148to110.090}). Like before, the parameters found by these fits are substituted in \Cref{disper_v04}, and this equation is solved to find the theoretical spectrum of the eigen-modes.

\begin{figure}[htbp]
\centering
\captionsetup[subfigure]{labelformat=empty}
\subcaptionbox{\label{Ekw_40.207to75.148}}{\includegraphics[width=1\linewidth]{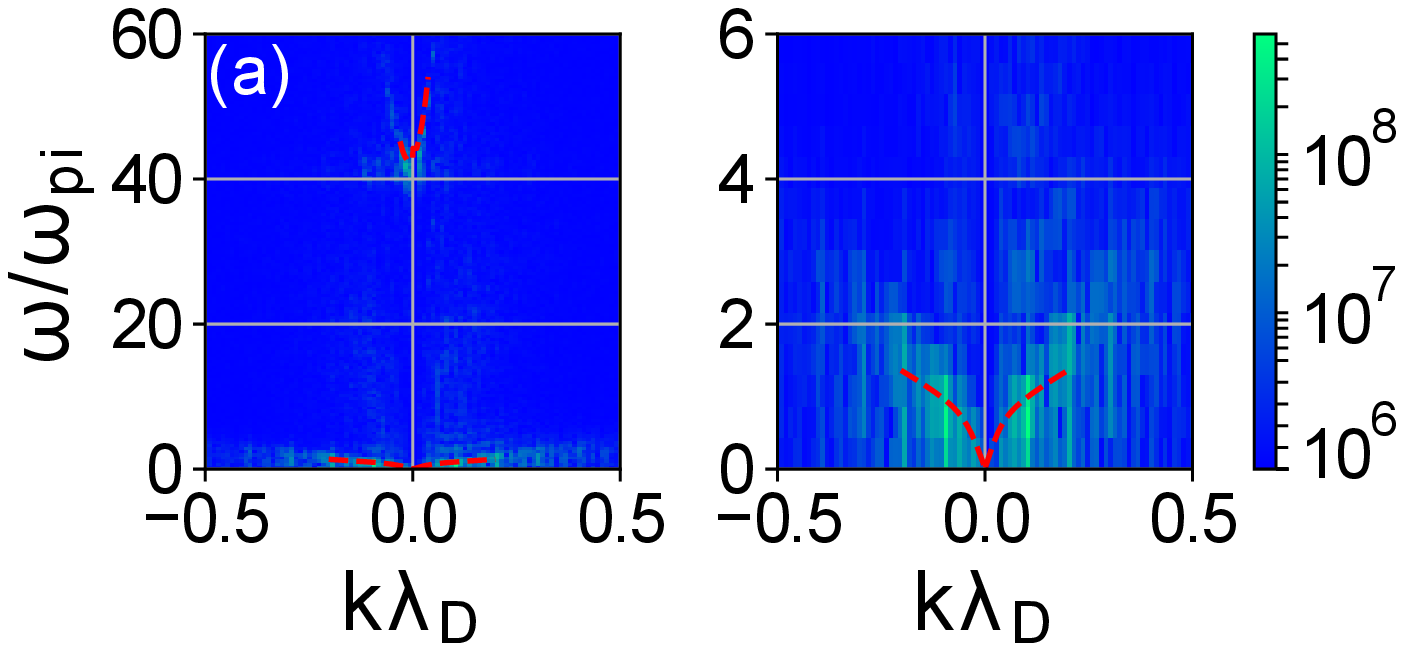}}\vspace{-2em}
\subcaptionbox{\label{Ekw_75.148to110.090}}{\includegraphics[width=1\linewidth]{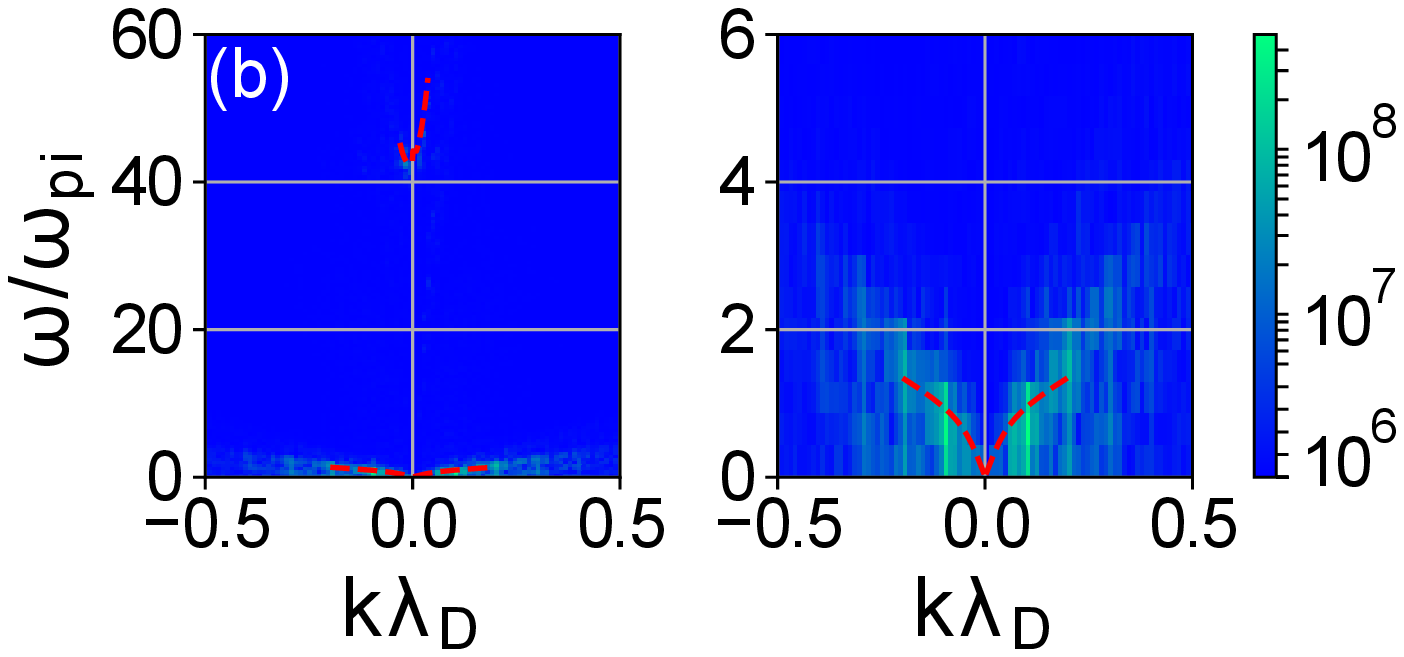}}\vspace{-2em}
\caption{The spectrum of nonlinear waves the case $v_0=10\,v_{te}$, for (a) $16.8\,\omega_{pi}^{-1}$ to $31.4\,\omega_{pi}^{-1}$ and (b) $31.4\,\omega_{pi}^{-1}$ to $46\,\omega_{pi}^{-1}$. The red lines show the modes found by solving \Cref{disper_v04}. In each case, a zoom into the low-frequency region is shown on the right of the full spectrum.}
\label{Ekw_v010}
\end{figure}

\begin{figure}[htbp]
\centering
\captionsetup[subfigure]{labelformat=empty}
\subcaptionbox{\label{disp_40.207to75.148}}{\includegraphics[width=.46\linewidth]{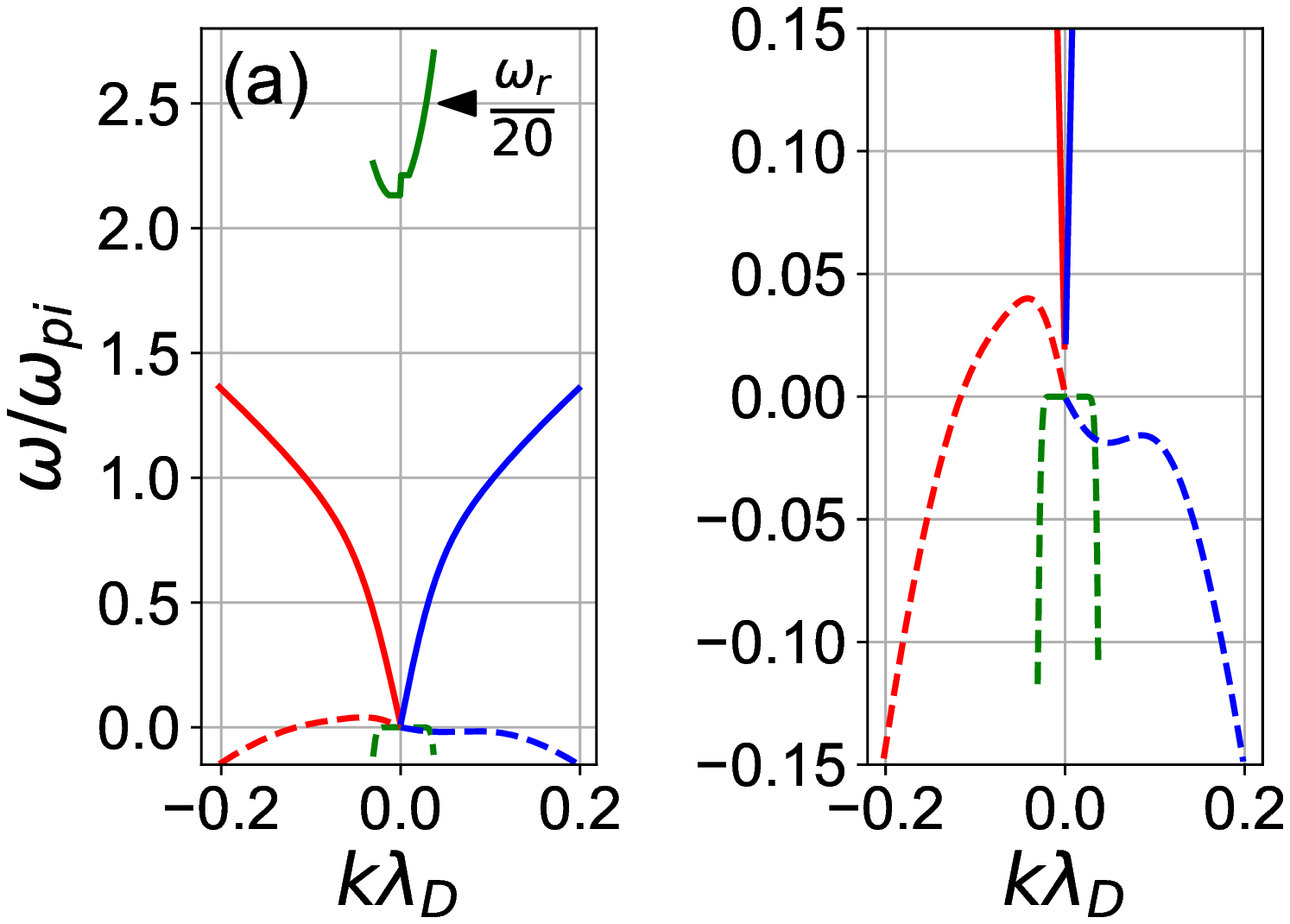}}
\subcaptionbox{\label{disp_75.148to110.090}}{\includegraphics[width=.46\linewidth]{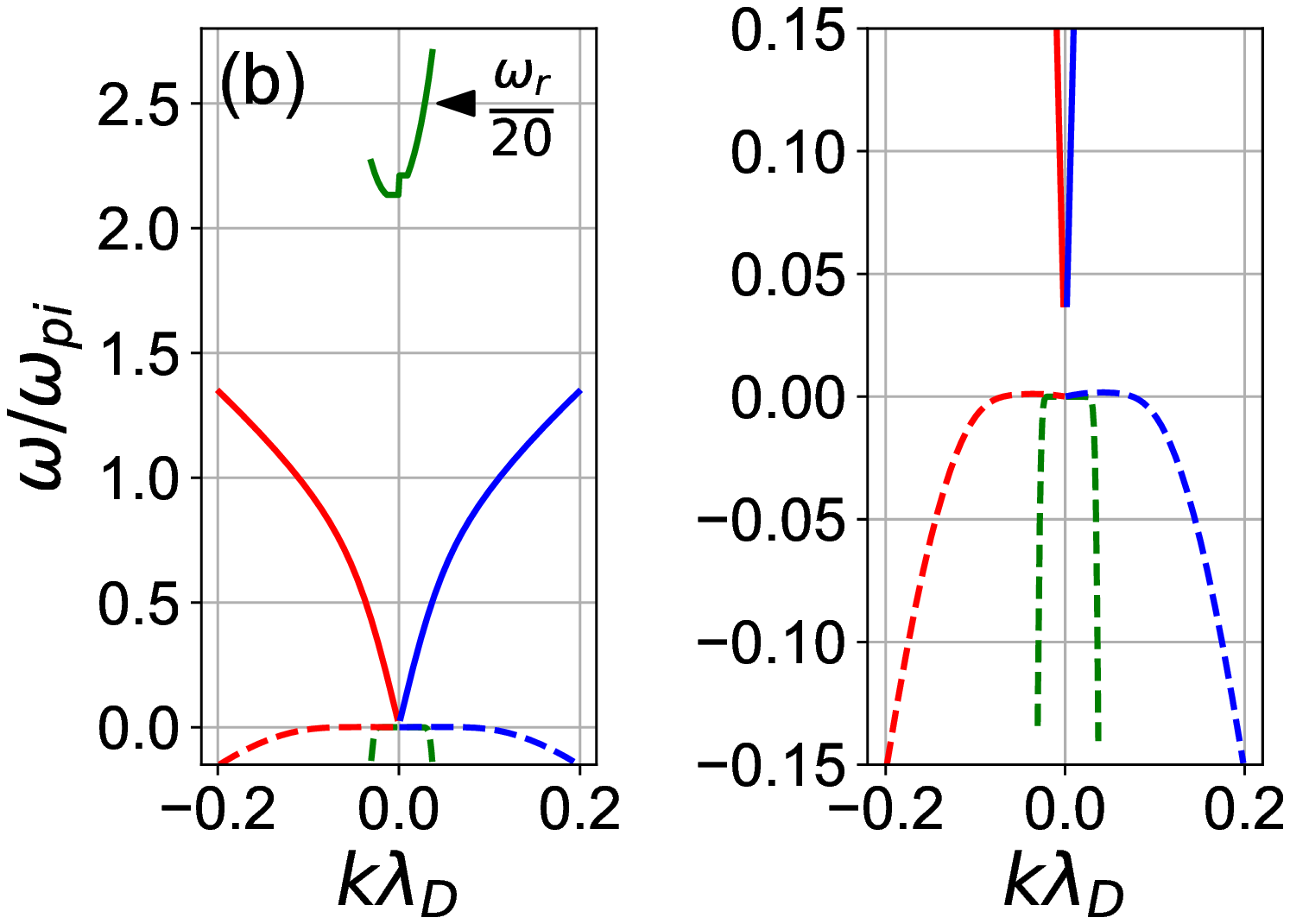}}
\caption{The theoretical frequencies (solid lines) and the growth rates (dashed lines) in two stages for the case $v_0=10\,v_{te}. $ (a) $16.8\,\omega_{pi}^{-1}$ to $31.4\,\omega_{pi}^{-1}$ and (b) $31.4\,\omega_{pi}^{-1}$ to $46\,\omega_{pi}^{-1}$. 
These results are found by solving \Cref{disper_v04}. The modes with growth rates less than $-0.15 \omega_{pi}$ are omitted. The real and imaginary parts corresponding to a given complex root are shown with the same color.}
\label{disp_v010}
\end{figure}

In the spectrum of the nonlinear waves, we see that the frequencies of the dominant modes (and therefore their phase velocities) are generally higher than the case of $v_0=4\,v_{te}$ (compare \Cref{Ekw_v04} with \Cref{Ekw_v010}). This feature is well captured by the theoretical model (compare \Cref{disp_v04} with \Cref{disp_v010}). Another difference between this case and that of $v_0=4v_{te}$ is the shorter range of the spectrum in the $k$ values. In the current case, we see that the modes with $\abs{k}\gtrsim0.2$ (as in contrast to $\abs{k}\gtrsim0.4$) are faint. This is consistent with the linear theory, which also shows a shorter unstable spectrum for the case of $v_0=10v_{te}$ (\Cref{linear_4vte,linear_10vte}). Like before, we see that the high-frequency modes fade way from the spectrum as time passes (see \Cref{Ekw_75.148to110.090}); this observation can be explained by the theoretical growth rates being negative in each time window (\Cref{disp_40.207to75.148}, \Cref{disp_75.148to110.090}).

\section{Drift velocity threshold for the appearance of backward waves}\label{Threshold}
\begin{figure}[htbp]
\centering
\captionsetup[subfigure]{labelformat=empty}
\subcaptionbox{\label{t_efE_v01.5}}{\includegraphics[width=.46\linewidth]{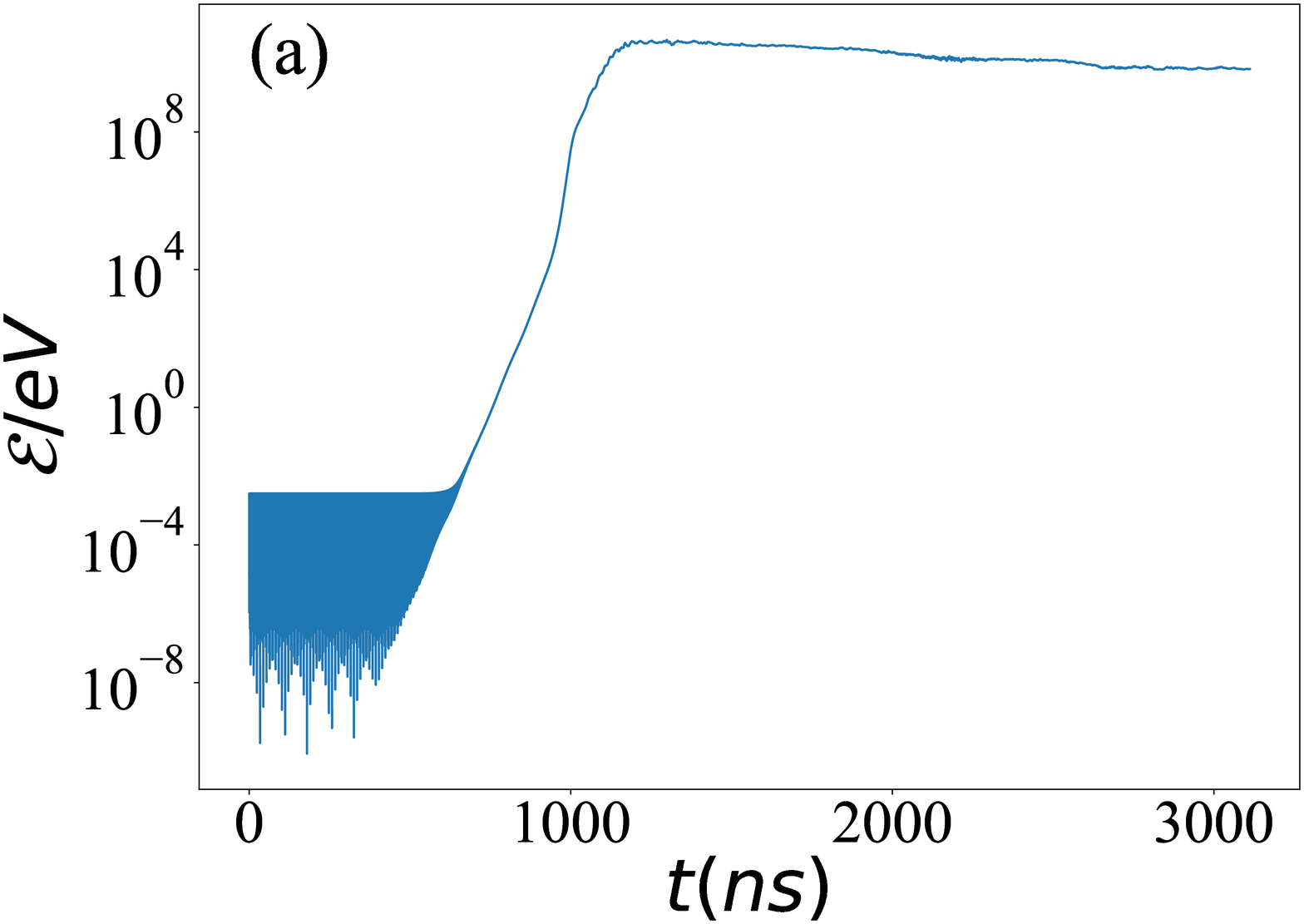}}
\subcaptionbox{\label{t_efE_v01.75}}{\includegraphics[width=.46\linewidth]{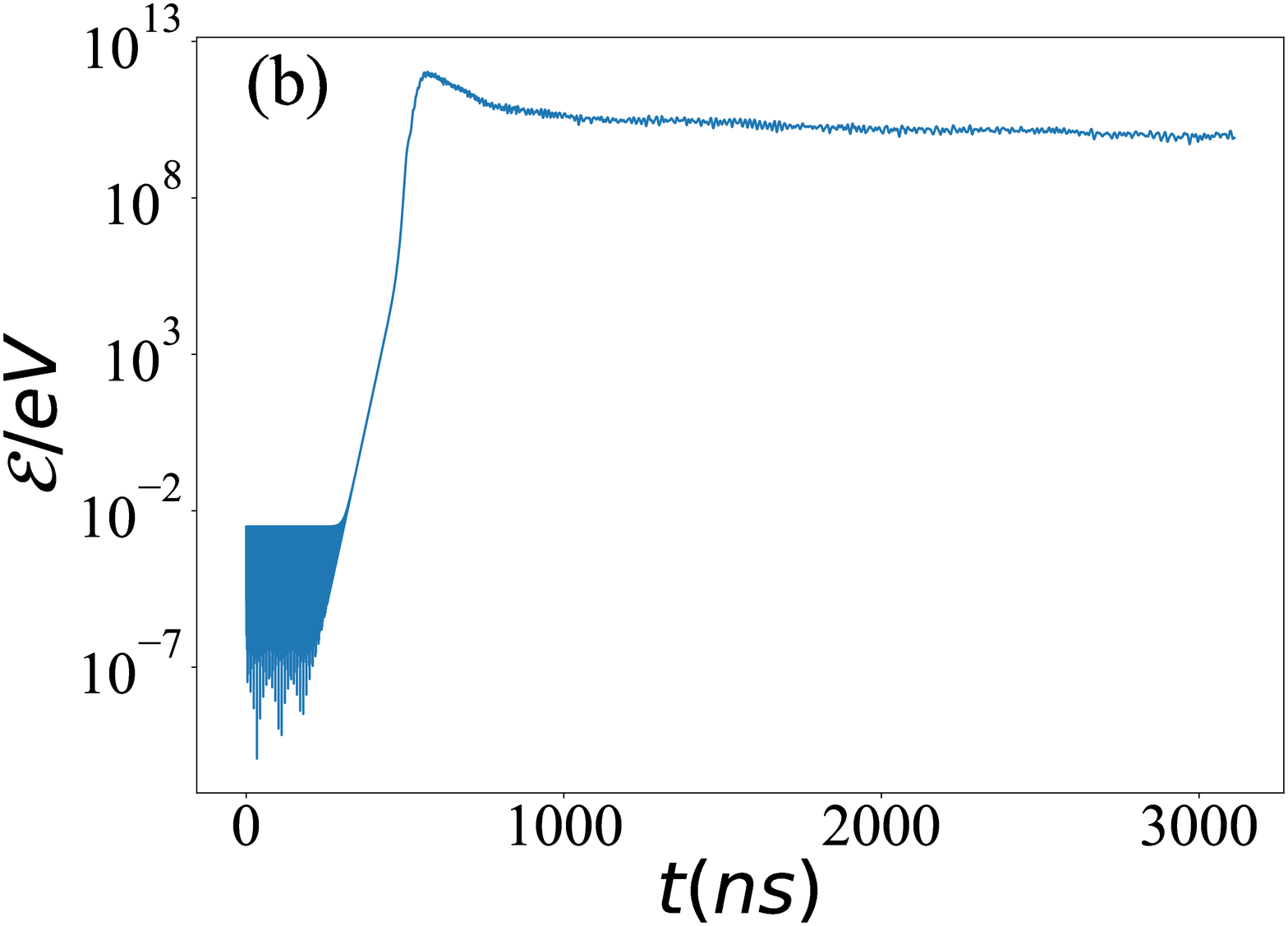}}
\caption{The electric field energy in two cases (a) $v_0=1.5v_{te}$ and (b) $v_0=1.75v_{te}$.}

\end{figure}

The backward waves do not appear if the drift velocity is below of some finite threshold $v_{tr}$. In order to narrow the down the  threshold value, we compare two simulations with $v_0=1.5v_{te}$ and $v_0=1.75v_{te}$. In both cases, all other parameters are unchanged. We show that the backward waves are not present in the former case, whereas they are present in the latter. Therefore, $v_{tr}$ is somewhere between these two values. \Cref{t_efE_v01.5,t_efE_v01.75} show the electric field energy for the cases of $v_0=1.5v_{te}$ and $v_0=1.75v_{te}$. By comparing these figures with \Cref{t_efE_v04} and \Cref{t_efE_v010}, we see that by increasing $v_0$ from $1.5v_{te}$ to $10v_{te}$, the saturation energy increases \cite{rajawat2017particle,HaraPSST2019}. However, because the linear growth rate also increases with $v_0$, the nonlinear regime is reached sooner. 

\begin{figure}[htbp]
\centering
\captionsetup[subfigure]{labelformat=empty}
\subcaptionbox{\label{total_ef_v01.5}}{\includegraphics[width=.46\linewidth]{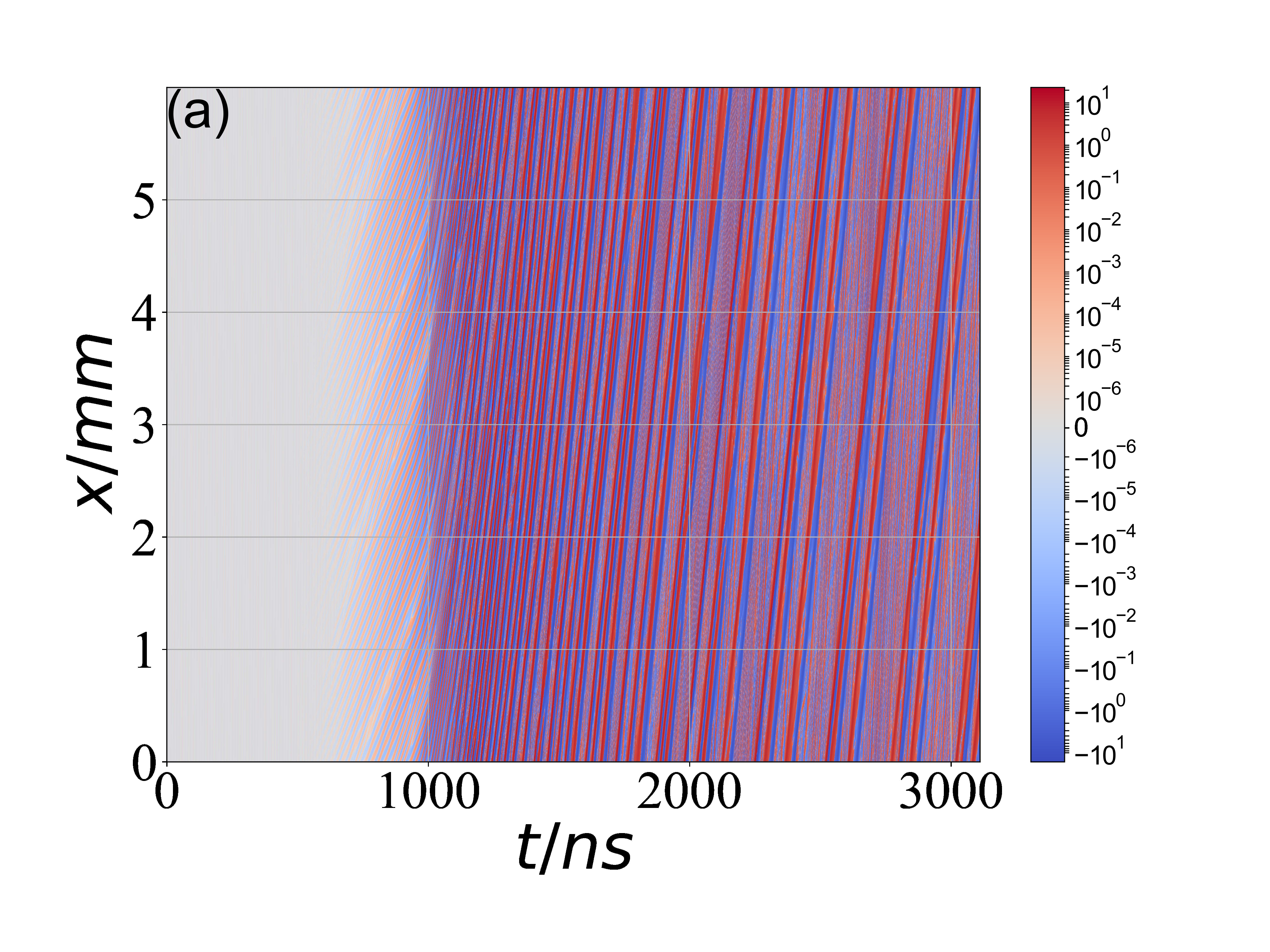}}
\subcaptionbox{\label{total_ef_v01.75}}{\includegraphics[width=.46\linewidth]{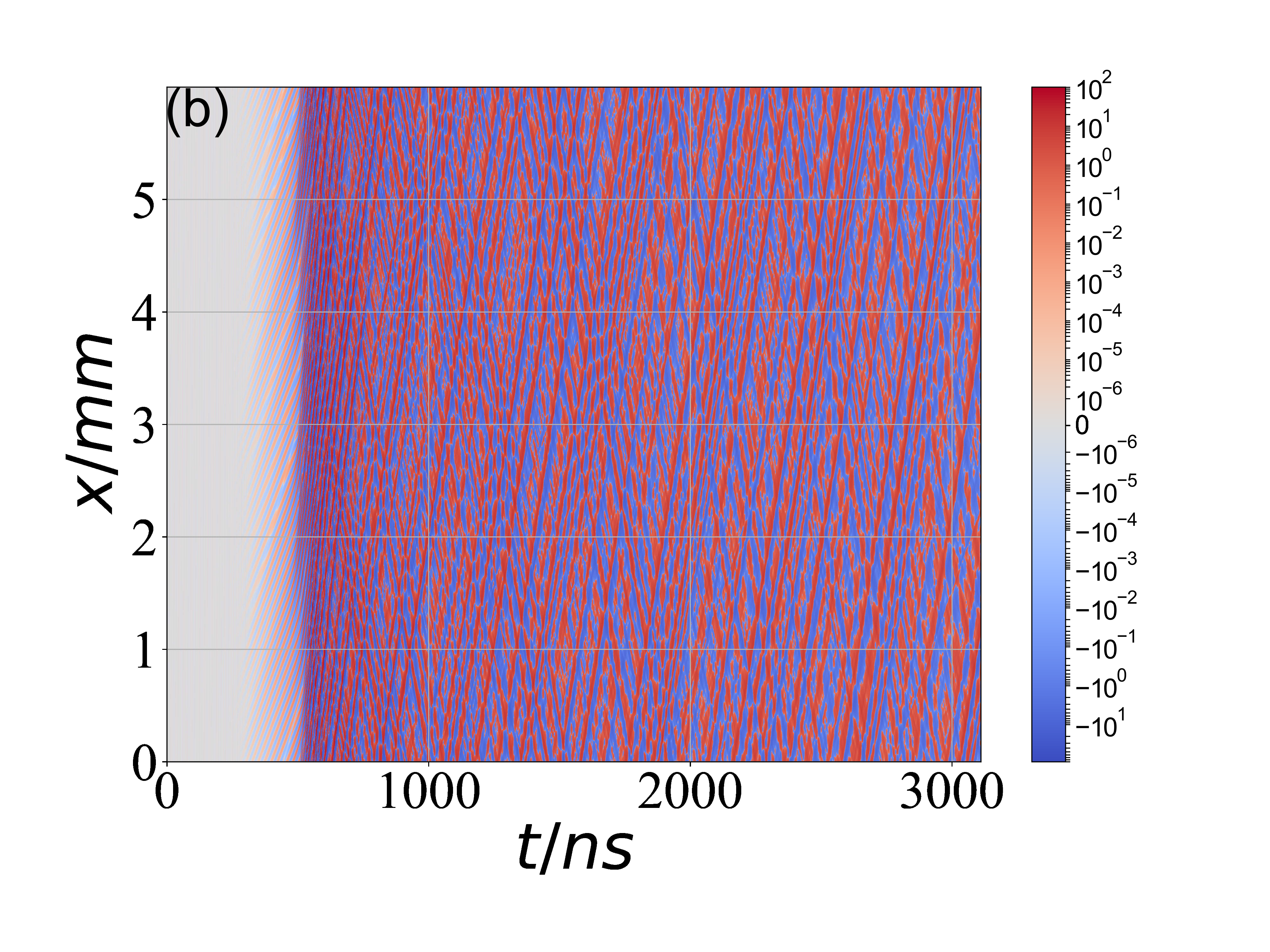}}
\caption{The electric field as a function of time and position in the cases of (a) $v_0=1.5v_{te}$ and (b) $v_0=1.75v_{te}$.}
\end{figure}

\begin{figure}[htbp]
\centering
\captionsetup[subfigure]{labelformat=empty}
\subcaptionbox{\label{fe2872_v01.5}}{\includegraphics[width=.46\linewidth]{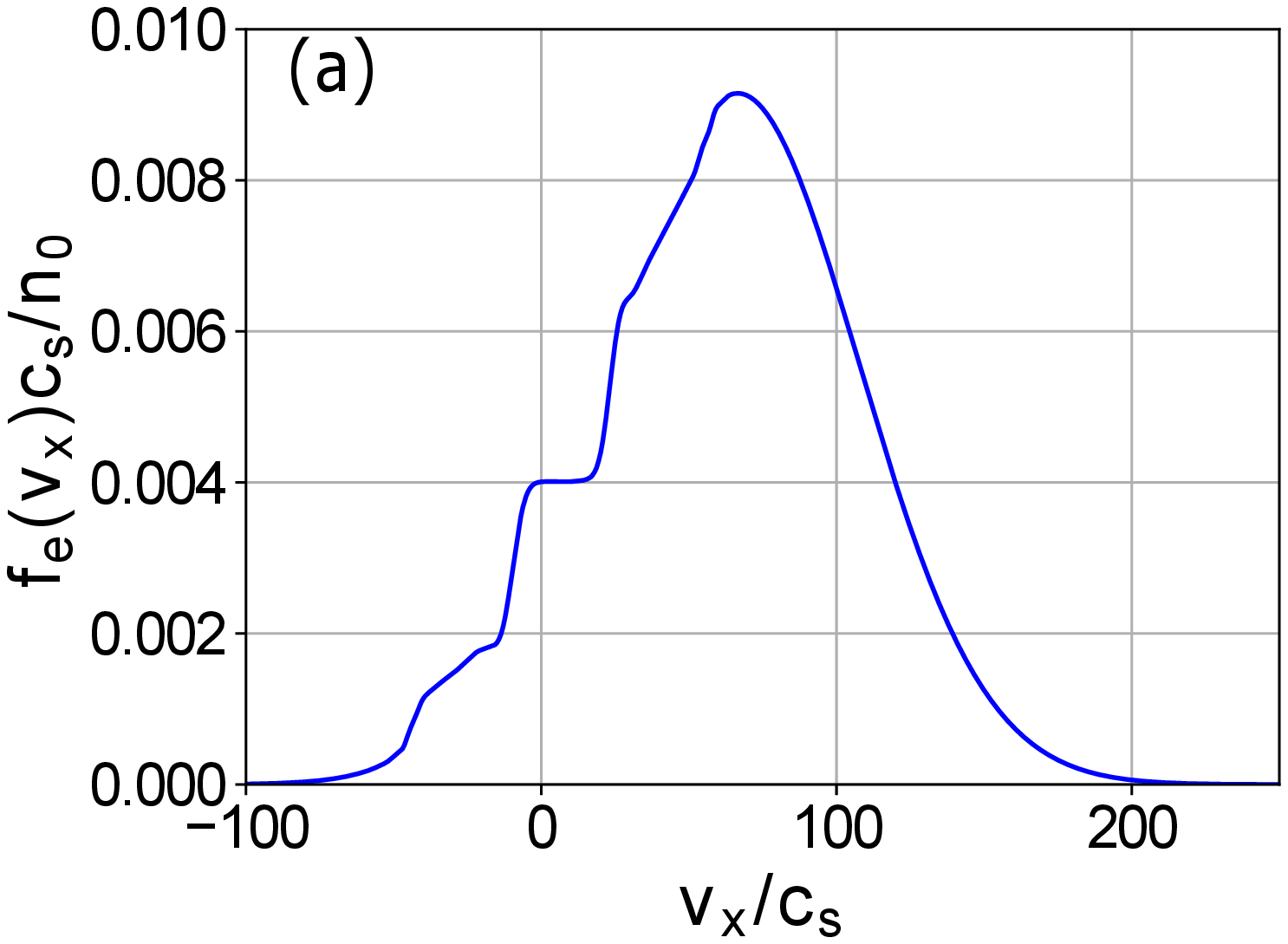}}
\subcaptionbox{\label{fe2872_v01.75}}{\includegraphics[width=.46\linewidth]{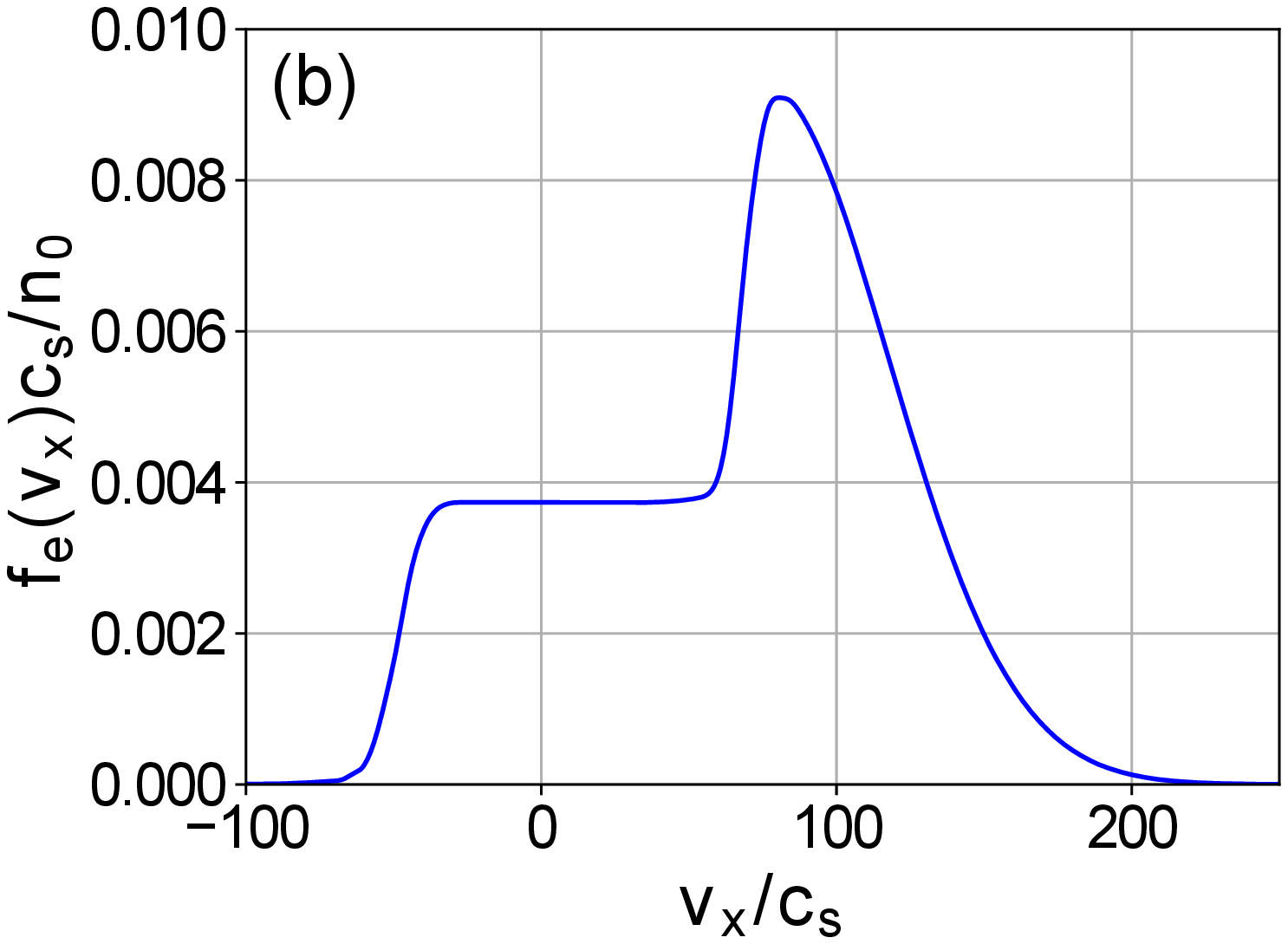}}
\caption{The electron VDF at time 2872 ns in the cases of (a) $v_0=1.5v_{te}$ and (b) $v_0=1.75v_{te}$.}
\end{figure}

\begin{figure}[htbp]
\centering
\captionsetup[subfigure]{labelformat=empty}
\subcaptionbox{\label{Ekw_2800to3000_v01.5}}{\includegraphics[width=.46\linewidth]{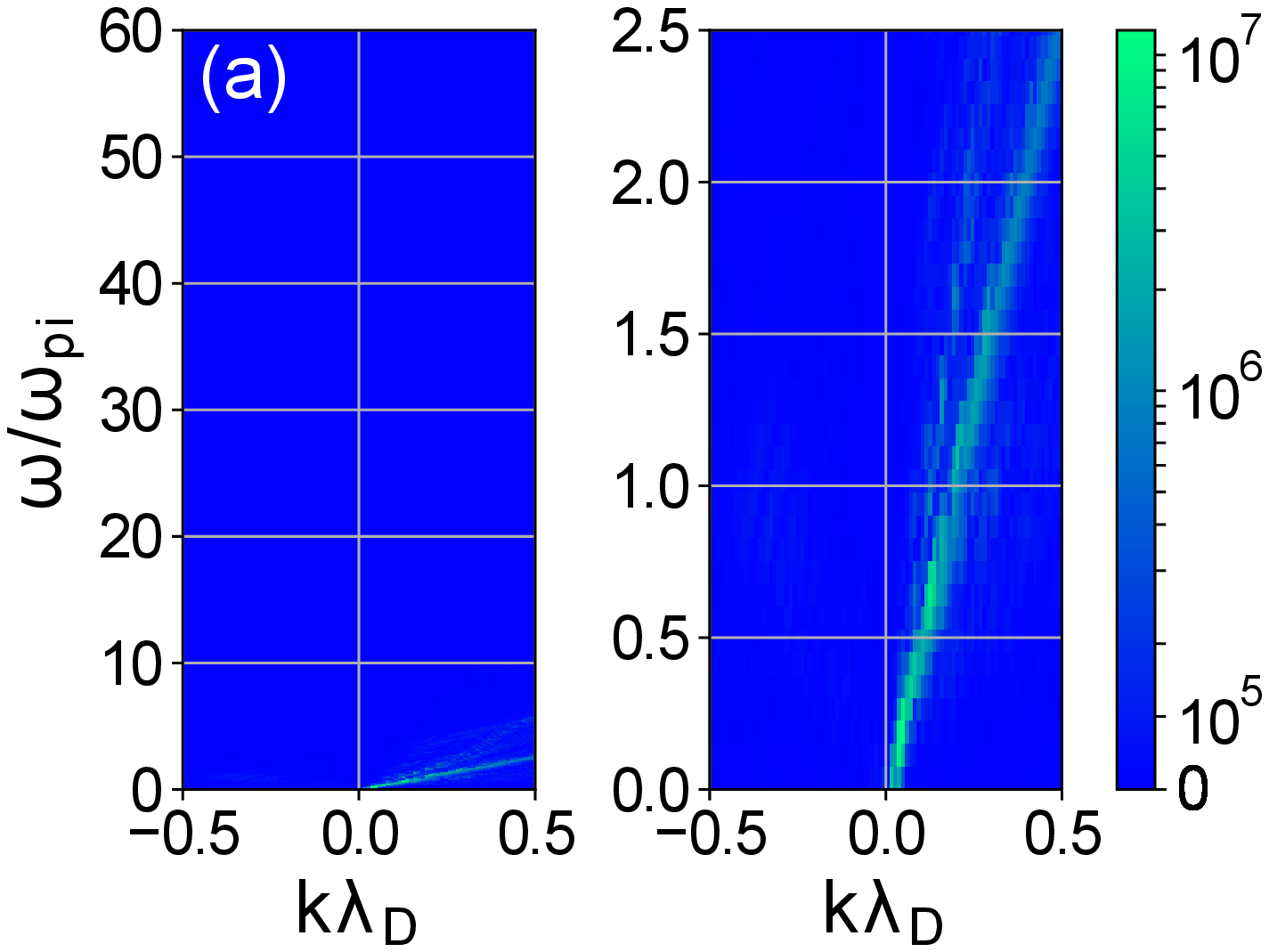}}
\subcaptionbox{\label{Ekw_2800to3000_v01.75}}{\includegraphics[width=.46\linewidth]{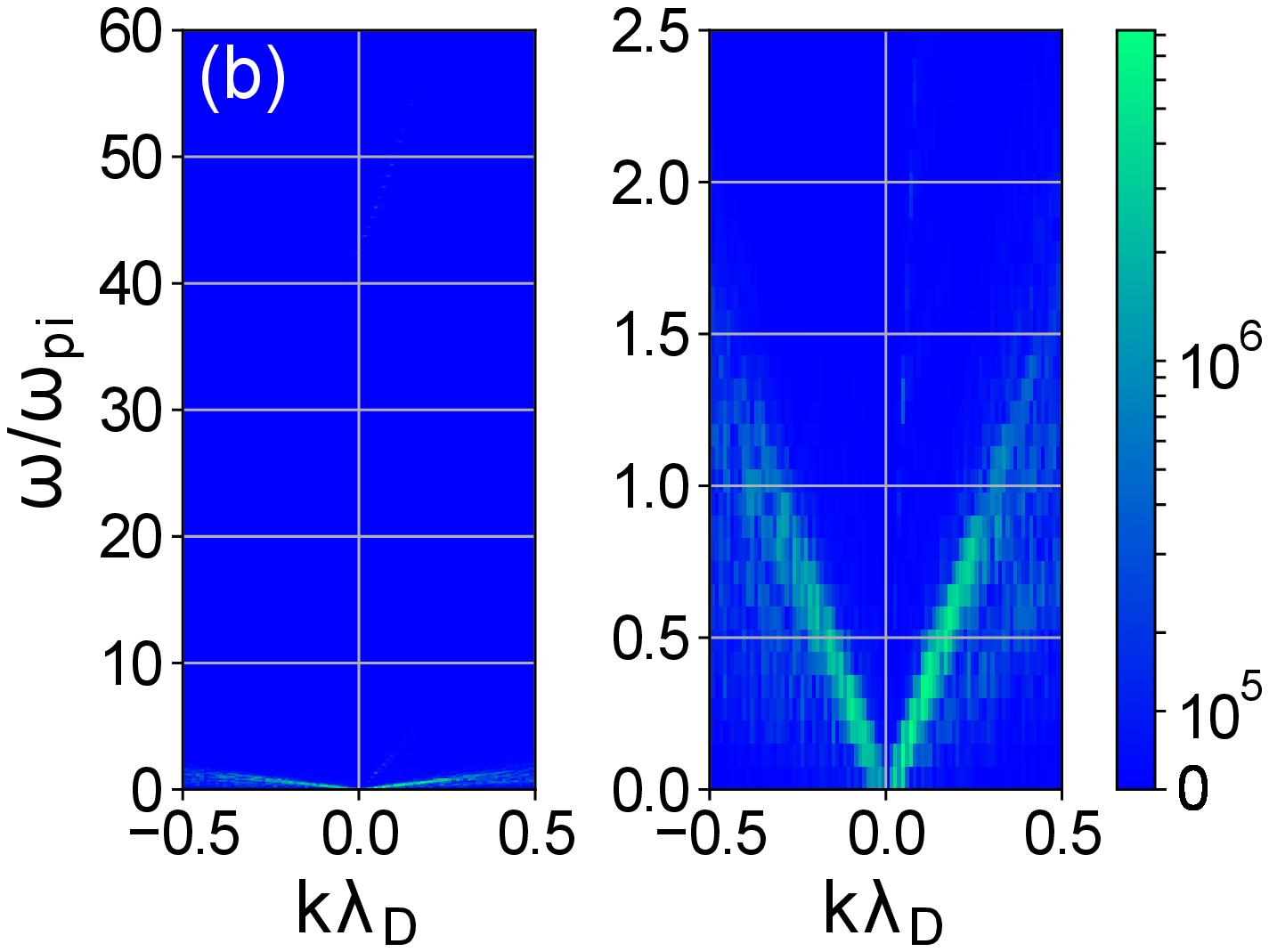}}
\caption{The Fourier modes between 2800 ns to 3000 ns in the cases of (a) $v_0=1.5v_{te}$ and (b) $v_0=1.75v_{te}$. In each case, a zoom into the low-frequency region is shown on the right of the the full spectrum.}
\end{figure}

\Cref{total_ef_v01.5} shows the electric field evolution in the case of $v_0=1.5v_{te}$. We see that, in this case, no backward waves appear, even deep in the  nonlinear regime. However, in the case of $v_0=1.75v_{te}$ (\Cref{total_ef_v01.75}) after about $600$ ns,  lines of negative slope appear, indicating the presence of backward waves. The explanation can again be given according to the linear theory of Landau damping using the nonlinear electron VDF. The electron VDF for the case $v_0=1.5v_{te}$ after equilibrium, (\Cref{fe2872_v01.5}), shows a plateau in the positive velocity region up to about $v_x=20\;c_s$. However, in the negative velocities, despite some nonlinear modifications, the gradient remains mainly positive. In fact, the trapping in this case is not strong enough to extend the plateau to the negative velocity region. Therefore, the absence  of the plateau leads to the lack of the backward waves in this case. In contrast, the plateau of the case $v_0=1.75v_{te}$ (\Cref{fe2872_v01.75}) is extended well into the region of negative velocities, and therefore, the marginally stable backward waves appear in this case. The absence of backward waves in the case of $v_0=1.5v_{te}$ and their excitation in the case of $v_0=1.75v_{te}$ can also be seen in the Fourier spectrum (\Cref{Ekw_2800to3000_v01.5,Ekw_2800to3000_v01.75}). We note that the time intervals of these spectra correspond to the end of the simulation, and therefore, the high-frequency modes have already disappeared, as in the previous cases.

The minimum required drift velocity for the excitation of backward waves found in our simulations with hydrogen  $v_0\sim 1.5v_{te}-1.75v_{te}$ is  above  the linear instability criteria  $v_0>1.3v_{te}$ for $T_i=T_e$ \cite{buneman1959dissipation}. In Ref.~\onlinecite{HaraPSST2019}, the threshold  $v_0=1.3v_{te}$ for the excitation of backward waves in plasmas with $T_e\gg T_i$ and heavier ions was reported.  We note, however, that the linear instability threshold is a function of the mass  and the temperature ratios,  and it is greatly reduced for lower values of $m_e/m_i$ and   $T_i\gg T_e$ \cite{buneman1959dissipation,JacksonPR1960,Mikhailovskii_v1} For $T_e\gg T_i$, the linear instability becomes the ion-sound like one and has a much lower  threshold $v_0> \mathcal{O}  (c_s)$~\cite{Mikhailovskii_v1}. 
 We have  confirmed in additional simulations (not reported here) that the threshold for the generation of  backward waves   also decreases with increase in ion mass and decrease in ion temperature with respect to the electrons.   


\section{Summary and Discussion}\label{discussion}

In this work, we investigated the backward waves that are excited in the nonlinear regime of the Buneman instability.  We have shown that the backward waves are  excited if the value of the initial drift velocity exceeds a certain threshold, which was found in  our simulations to be in the range between $v_0=1.5v_{te}$ to $v_0=1.75v_{te}$.  
 Using the  dispersion stability analysis, we have shown that  characteristics of the backward and forward waves observed in nonlinear simulations can be explained,  both in  high-frequency and low-frequency regions, as marginally stable configurations  of the nonlinear electron distribution function. We have found that an  extended plateau in the region of the phase velocity of backward waves (in the region of the negative electron velocities) is necessary for the formation of the backward  waves. 
This is further confirmed  by the simulations with $v_0=1.5v_{te}$, where the absence of the plateau in the negative velocity region of electron VDF prevents the excitation of backward waves. 

In a similar approach, the formation of a plateau in the ion VDF (in Ref.~\onlinecite{valentini2011new}) and the electron VDF (in Refs.~\onlinecite{shoucri2017formation,valentini2006excitation,valentini2012undamped}) has been found responsible for the new class of waves not expected from the linear theory due to finite Landau damping. In those studies, however, the modes were generally of smaller amplitude, and the contribution of the trapped particles was limited to a short range of velocities inside a narrow plateau in the VDF. In our simulations, the backward waves 
are generated simultaneously with formation of the electron beams in the negative direction due to electron reflections from large-amplitude potential structures as illustrated in \Cref{x_phi} of the spatial potential profile. This process,  starting with the electron trapping into the electron holes,
proceeds via the growth of the potential fluctuations  and further electron scattering (due to trapping and de-trapping). As a result of such scattering, the electrons are heated with the formation of the electron VDF with an extended plateau in the region of the negative velocities.    The resulting potential structures have large  amplitudes, and the plateau in the electron VDF  covers a significant region of the velocity range (i.e.,~many of the electrons are trapped). Our theoretical analysis shows that the nonlinearly modified distribution functions represent the marginally stable configuration from the perspective of linear stability.

In Ref.~\onlinecite{HaraPSST2019}, the ion kinetic effects associated with ion heating and backward waves generated in the nonlinear regime of current-driven instabilities were studied, in particular, as relevant to the hollow-cathode discharges and cathode surface  sputtering. We note that in our study, the ions are not heated for $v_0=4v_{te}$ and only weakly heated for $v_0=10v_{te}$.   Backward waves have also been reported in the particle-in-cell simulation of the Buneman instability in Ref.~\onlinecite{jun2012competition}. Excitation of the backward waves was observed to co-exist with an enhanced anomalous resistivity in Ref.~\onlinecite{DyrudJGR2006}. Similarly, backward waves were also suggested as the reason for an increase in the effective collision frequency seen in the simulation of Ref.~\onlinecite{HellingerGRL2004}. In that work, it was conjectured that the underlying mechanism for the generation of the backward waves is the induced scattering off the ions. The resonant condition for the induced scattering from ions have the form $\omega(k_1)-\omega(k_2)\sim (k_1-k_2)v_{ti}$.
 The low-frequency modes in our simulations show a symmetry between the forward and backward spectra (i.e., $\omega(k_1)\approx\omega(k_2)$ for these modes) and $k=k_1=-k_2$. Moreover, for these low-frequency modes, one has $\omega(k_1)\approx\omega(k_2) \sim \abs{k v_{ti}}$.
 This resonant condition is difficult to satisfy
 for the symmetric modes when $\omega(k_1)-\omega(k_2) \ll 2 \abs{k v_{ti}}$.  Therefore, it is unlikely that in our case the induced scattering off the ions is responsible for the excitation of  the symmetric spectra of low-frequency backward and forward waves.   Based on nonlinear weak turbulence theory for the bump-on-tail instability,  Ref.~\onlinecite{YoonPoP2018} attributes the high-frequency backward waves to the combined effect of three-wave decay and scattering off the ions and the low-frequency backward waves to merely the three-wave-decay process. However, the low-frequency modes in Ref.~\onlinecite{YoonPoP2018} are transient and decay to the level of noise later in the nonlinear regime. In Ref.~\onlinecite{YoonPoP2018},   the formation of the plateau of the electron distribution function in the high-velocity region (of order $v_{te}$)  is consistent with the observed high-frequency backward waves.  As we have shown above  in our simulations, the  wide plateau extending into the negative region (and thus covering the low velocity region)  is responsible for the 
 sustainment of the  low-frequency ion-sound-like modes. The backward waves have been investigated in  Vlasov simulations~\cite{jain2011modeling} for the case of $v_0=4v_{te}$.  That study suggests the origin of the backward waves can be a secondary linear instability driven by bulks of counter-streaming electrons. However,   the analysis was focused on the high-frequency modes, and the FFT in time was not long enough to clearly show the dominant low-frequency modes. 
The Buneman-type instability, excited by electron beams from solar nanoflares, was proposed as an underlying mechanism for the generation of turbulence in solar wind, plasma heating, and anomalous resistivity \cite{CheHH_MPLA2016,ChePoP2017,HellingerGRL2004}. We show here that electron (plasma) heating and backward waves are closely related. It is also expected that anomalous resistivity is also affected. The effective heating of electrons observed in the regime $v_0 \geq v_{te} $  may be of interest for industrial applications of plasma-beam discharges.  It has been suggested \cite{HaraPSST2019} that backward  waves may be related  to the hollow cathode erosion in Hall thruster due to the subsequent acceleration of the ions. This process  was not considered in our study   but  potentially may occur at  a later stage and with non-periodic setup when there is a constant energy input to the system.  Intriguing  experimental observations of the waves propagating against the direction of the electron beam were reported in Ref.~\onlinecite{TsikataPoP2009}. No theoretical explanation has been proposed so far. It is possible that such waves are excited via the mechanism considered in our study.

Two-dimensional effects can be important and can  modify the nonlinear dynamics of trapping and heating \cite{hutchinson2017electron}. It is however expected that in  applications to magnetized plasma,  the transverse plasma motion is constrained  by the magnetic field, whereas the dynamics along the magnetic field is expected to be well approximated by the one-dimensional model, as used in this study and other previous works\cite{che2009nonlinear,che2010electron}.
Our results show that the backward waves are related to particle trapping and accompanied by the formation of the electron holes, which extend the plateau in the electron VDF to the negative velocity region. The dynamics and conditions of electron holes in multi-dimensional plasma is a multifaceted  topic, where many questions remain open \cite{hutchinson2017electron}. Some computational studies show that the stable electron holes do not appear in isotropic multi-dimensional plasmas \cite{miyake1998two,oppenheim2001evolution,lu2008perpendicular,hutchinson2017electron}. The reason is that the fraction of trapped particles is proportional to $\phi_0^{D/2}$, where $\phi_0$ is the amplitude of potential and $D$ is the number of spatial dimensions. Therefore for $D=2$ or $D=3$, the population of trapped particles may be too small  to furnish  stable electron holes \cite{krasovsky2004effect}.   However, strong magnetic fields can make the dynamics quasi-one-dimensional along the magnetic field,  and  stable electron holes can appear \cite{miyake1998two, lu2008perpendicular}. In such a situation,  the parallel electron VDF may form a plateau extending to the negative velocity region and leading to  onset of the backward waves as discussed above. The situation may become more complex when  narrowly localized electron beams with a width of the order of the electron skin-depth across the magnetic field are involved,  and two-dimensional effects may become essential, e.g., leading to the transverse ion heating Ref.~\onlinecite{reitzel1998dynamics}.
Nevertheless, it is interesting to note, that  even in this case, the two-dimensional simulations Ref.~\onlinecite{reitzel1998dynamics} demonstrate the electron heating and show the electron VDF extending to the negative velocity region, similar to the case considered here. The existence of  backward waves in multi-dimensional situations requires additional analysis that is beyond the scope of this study. The electron trapping and hole formation may also be prevented  by collisions when  the collision frequency exceeds the characteristic bounce frequency for trapped electrons.  

\section{Acknowledgment}
This work is partially supported in part by  US Air Force Office of Scientific Research FA9550-15-1-0226, the Natural Sciences and  Engineering Council of Canada (NSERC), and Compute Canada computational resources.  
 \section{Data Availability Statement}
 The data that support the findings of this study are available from the corresponding author
upon reasonable request.

\bibliographystyle{apsrev4-2}
\bibliography{Refs.bib}

\end{document}